\documentclass[oldversion]{aa}

\usepackage[english]{babel}
\usepackage[dvips]{graphicx}
\usepackage{latexsym}
\usepackage{amssymb}
\usepackage{mathrsfs}
\usepackage{multirow}
\usepackage{ctable}
\usepackage{url}
\usepackage[authoryear]{natbib}
\bibpunct{(}{)}{;}{a}{}{,}

\begin{document}

\newcommand{\hours}{^{\mathrm{h}}}
\newcommand{\mins}{^{\mathrm{m}}}
\newcommand{\secs}{^{\mathrm{s}}}
\newcommand{\degs}{^{\circ}}
\newcommand{\jybeam}{Jy~beam$^\mathrm{-1}${}}
\newcommand{\mjybeam}{mJy~beam$^\mathrm{-1}${}}
\newcommand{\ra}{\mathrm{RA}}
\newcommand{\dec}{\mathrm{DEC}}
\newcommand{\bootes}{Bo\"otes}
\newcommand{\solarmass}{M$^{}_{\odot}$}
\newcommand{\micron}{$\mu$m}
\newcommand{\fri}{FR{\sc{}i}}
\newcommand{\frii}{FR{\sc{}ii}}

\title{Deep Low-Frequency Radio Observations of the \\ NOAO Bo\"otes Field}
\subtitle{I. Data Reduction and Catalog Construction}
\titlerunning{Deep Low-Frequency Radio Observations of the Bo\"otes Field I.}

\author{H.~T.~Intema\inst{1,2}\thanks{Jansky Fellow of the National Radio Astronomy Observatory}
   \and R.~J.~van~Weeren\inst{1}
   \and H.~J.~A.~R\"ottgering\inst{1}
   \and D.~V.~Lal\inst{3,4}}

\authorrunning{H.~T.~Intema et al.}
\offprints{\email{hintema@nrao.edu}}

\institute{Leiden Observatory, Leiden University, P.O.Box 9513, NL-2300 RA, Leiden, The Netherlands
      \and National Radio Astronomy Observatory, Charlottesville, VA, USA
      \and National Centre for Radio Astrophysics, Univerity of Pune, India
      \and Max-Planck-Institut f\"ur Radioastronomie, Bonn, Germany}

\date{Received ... / Accepted ...}

\abstract{In this article we present deep, high-resolution radio interferometric observations at 153~MHz to complement the extensively studied NOAO \bootes{} field. We provide a description of the observations, data reduction and source catalog construction. From our single pointing GMRT observation of $\sim 12$~hours we obtain a high-resolution ($26\arcsec \times 22\arcsec$) image of $\sim 11.3$~square degrees, fully covering the \bootes{} field region and beyond. The image has a central noise level of $\sim 1.0$~\mjybeam{}, which rises to 2.0--2.5~\mjybeam{} at the field edge, placing it amongst the deepest $\sim 150$~MHz surveys to date. The catalog of 598~extracted sources is estimated to be $\sim 92$~percent complete for $> 10$~mJy sources, while the estimated contamination with false detections is $< 1$~percent. The low RMS position uncertainty of $1.24\arcsec$ facilitates accurate matching against catalogs at optical, infrared and other wavelengths. Differential source counts are determined down to $\lesssim 10$~mJy. There is no evidence for flattening of the counts towards lower flux densities as observed in deep radio surveys at higher frequencies, suggesting that our catalog is dominated by the classical radio-loud AGN population that explains the counts at higher flux densities. Combination with available deep 1.4~GHz observations yields an accurate determination of spectral indices for 417~sources down to the lowest 153~MHz flux densities, of which 16 have ultra-steep spectra with spectral indices below $-1.3$. We confirm that flattening of the median spectral index towards low flux densities also occurs at this frequency. The detection fraction of the radio sources in NIR $K_S$-band is found to drop with radio spectral index, which is in agreement with the known correlation between spectral index and redshift for brighter radio sources.}


\maketitle

\section{Introduction}
\label{sec:bootes_intro}

Surveying the radio sky at low frequencies ($\lesssim 300$~MHz) is a unique tool for investigating many questions related to the formation and evolution of massive galaxies, quasars and clusters of galaxies \citep[e.g.,][]{miley2008}. Low-frequency radio observations benefit from the steepness of radio spectra of various types of cosmic radio sources, such as massive HzRGs (high-redshift radio galaxies; redshift $z \gtrsim 2$) and diffuse halo \& relic emission in nearby galaxy clusters ($z \lesssim 0.1$).

HzRGs are amongst the most massive galaxies in the early Universe \citep[e.g.,][]{miley2008}, usually located in forming galaxy clusters with total masses of more than $10^{14}_{}$~\solarmass{} \citep[e.g.,][]{venemans2007}. The most efficient way of finding HzRGs is to focus on USS (ultra-steep spectrum; $S_{\nu} \propto \nu^{\alpha}$ with $\alpha \lesssim -1$) radio sources \citep{rottgering1997,debreuck2002}. This was reinforced by \citet{klamer2006} who showed that the radio spectra of HzRGs in general do not show spectral curvature, but are straight. The USS selection criteria appear to hold down to very low flux levels \citep[e.g.,][]{afonso2011}. Concentrating on the faintest sources from surveys made at the lowest frequencies is therefore an obvious way of pushing the distance limit for HzRGs beyond the present highest redshift of TN~J0924-2201 at $z = 5.1$ \citep{vanbreugel1999} and probing massive galaxy formation into the epoch of reionization. 

Galaxy clusters containing diffuse radio sources appear to have large X-ray luminosities and galaxy velocity dispersions \citep[e.g.,][]{hanisch1982}, which are thought to be characteristics of cluster merger activity \citep[e.g.,][]{giovannini2000,kempner2001}. Synchrotron halos and relics provide unique diagnostics for studying the magnetic field, plasma distribution and gas motions within clusters, important inputs to models of cluster evolution \citep[e.g.,][]{feretti2004}. Cluster synchrotron emission is known to be related to the X-ray gas and pinpoints shocks in the gas. Further, cluster radio emission usually has steep radio spectra ($\alpha < -1$), the radiating electrons are old and can provide fossil records of the cluster history \citep[e.g.,][]{miley1980}.

Several low-frequency surveys have been performed in the past, such as the Cambridge surveys 3C, 4C, 6C and 7C at 159, 178, 151 and again 151~MHz, respectively \citep{edge1959,bennett1962,pilkington1965,gower1967,hales1988,hales2007}, the UTR-2 sky survey between 10-25~MHz \citep{braude2002}, the VLSS at 74~MHz \citep{cohen2007} and the ongoing MRT sky survey at 151.5~MHz \citep[e.g.,][]{pandey2007}. These surveys are limited in sensitivity and angular resolution, mainly due to man-made RFI, ionospheric phase distortions and wide-field imaging problems. Recent developments in data reduction techniques make it possible to perform deeper surveys ($\lesssim 50$~\mjybeam{}) of the low-frequency sky at higher resolution ($\lesssim 30\arcsec$). One telescope that significantly improved this situation is the GMRT. A few deep, single-pointing surveys at 153~MHz \citep[e.g.,][]{ishwara2007,sirothia2009,ishwara2010} have been performed, yielding noise levels between 0.7-2~\mjybeam{} and resolutions between 20-30\arcsec{}. And the TGSS\footnote{\url{http://tgss.ncra.tifr.res.in/}} is a new, ongoing 153~MHz GMRT sky survey, aimed at covering the full northern sky down to $\dec{} > -30\degs$ at a $\sim 20\arcsec$ resolution and a $7-9$~\mjybeam{} noise.

In this article we present deep, high-resolution GMRT observations at 153~MHz of the NOAO \bootes{} field. The \bootes{} field is a large ($\sim 9$~square degree) northern field that has been targeted by surveys spanning the entire electromagnetic spectrum. This field has been extensively surveyed with radio telescopes including the WSRT at 1.4~GHz \citep{devries2002}, the VLA at 325~MHz \citep{croft2008}, and will be complemented with deep 74~MHz EVLA observations. The large northern NDWFS survey \citep{jannuzi1999} provided 6~colour images ($B^{}_{W} \, R \, I \, J \, H \, K$) to very faint optical and NIR flux limits. Additional, deeper NIR images in $J \, K^{}_{S}$ are available from the FLAMEX survey \citep{elston2006}, while additional $z$-band images are available from the $z$Bootes campaign \citep{cool2007}. The entire area has also been surveyed by Spitzer in seven IR bands ranging from 3.6 to 160~\micron{} \citep{eisenhardt2004,houck2005}. Chandra has covered this area in the energy range of 0.5--7~keV to a depth of $\sim 10^{-14}_{}$~ergs~s$^{-1}_{}$~cm$^{-2}_{}$, yielding 3200~quasars and 30~luminous X-ray clusters up to redshift $z \sim 1$ \citep{murray2005,kenter2005}. The UV space telescope GALEX has covered the \bootes{} field. All the 10,000 galaxies brighter than $R < 19.2$ and X-ray / IRAC / MIPS QSOs brighter than $R < 21.5$ have redshifts through the AGES project (Kochanek et al.~\textit{in preparation}). Based on the shallow Spitzer data, 3~HzRGs with photometric redshifts of $z > 4$ have been identified in the \bootes{} field \citep{croft2008}. Also, \citet{cool2006} report the discovery of 3~quasars with spectroscopic redshifts $z > 5$, while \citet{mcgreer2006} found a quasar at $z = 6.1$. 

Given the size of the GMRT FoV (field-of-view $\sim 3.5$~degrees) and angular resolution ($\sim 25\arcsec$) at 153~MHz, the \bootes{} field is a well-matched region for conducting a deep survey. Combined with the existing multi-wavelength surveys, our deep 153~MHz \bootes{} field observations allow for a complete study of faint ($\gtrsim 20$~mJy) low-frequency radio sources. For the data reduction, we used the recently developed SPAM calibration software that solves for spatially variant ionospheric phase rotations \citep[][]{intema2009a}.
In our initial analysis of the 153~MHz source catalog we focus on determining source counts down to the detection limit, and identifying steep-spectrum radio sources that are candidate HzRGs. A more extended analysis of our source catalog in combination with the other multi-wavelength catalogs is planned in a subsequent article.

In Section~\ref{sec:bootes_obs_dr}, we describe the GMRT 153~MHz observations and data reduction. In Section~\ref{sec:bootes_catalog}, we present details on the source extraction and catalog construction. Section~\ref{sec:bootes_analysis} contains the initial analysis and discussion of the source population. Conclusions are presented in Section~\ref{sec:bootes_concl}. Throughout this article, source positions are given in epoch J2000 coordinates.

\section{Observations and Data Reduction}
\label{sec:bootes_obs_dr}

In this section, we describe the observations and data reduction steps that led to the production of the 153~MHz image that is the basis of the survey.

\subsection{Observations}
\label{sec:bootes_obs}

We used the GMRT \citep[e.g.,][]{nityananda2003} at 153~MHz to observe the \bootes{} field, with the observing details given in Table~\ref{tab:bootes_observing}. There were typically 27 out of 30~antennas available during each observing run. We used 3C~286 as flux, bandpass and phase calibrator. Observation cycles of $\sim 50$ minutes were split into $\sim 38$~minutes on the target field and $\sim 10$ scans on the calibrator. The relatively high overhead in calibrator observations was justified by the need to monitor the GMRT system stability, RFI conditions and ionospheric conditions, and to ensure consistency of the flux scale over time.

\ctable[botcap,center,
caption = {Overview of GMRT observations on the \bootes{} field.},
label = tab:bootes_observing
]{l c c}{
\tnote[a]{Indian Standard Time (IST) = Universal Time (UT) + 5:30}
}{\FL
Date & June 3, 2005 & June 4, 2005 \ML
LST range & 12--20~hours & 10--19~hours \NN
Local time range\tmark[a] & 20:00--04:00 & 18:00--03:00 \NN
Time on target & 359~min & 397~min \NN
Primary calibrator & 3C~286 & 3C~286 \NN
Time on calibrator & 94~min & 108~min \ML
Integration time & \multicolumn{2}{c}{16.8~sec} \NN
Polarizations & \multicolumn{2}{c}{RR, LL} \NN
No. of channels & \multicolumn{2}{c}{128} \NN
Channel width & \multicolumn{2}{c}{62.5~KHz} \NN
Total bandwidth & \multicolumn{2}{c}{8.0~MHz} \NN
Target RA,DEC & \multicolumn{2}{c}{$14\hours32\mins05.75\secs$, $+34\degs16\arcmin47.5\arcsec$} \LL
}

\subsection{Data Reduction}
\label{sec:bootes_dr}

The data reduction was performed in two stages. The first stage consisted of `traditional' calibration, in which the flux scale, bandpass shapes and phase offsets were determined from the calibrator observations, transferred to the target field data, after which the target field was self-calibrated and imaged for several iterations. During the second stage of the data reduction, we made use of the recent SPAM software package \citep{intema2009a} that incorporates direction-dependent ionospheric phase calibration. 

For the traditional calibration we used the Astronomical Image Processing Sofware \citep[AIPS; e.g.,][]{bridle1994} package. Initial flagging of bad baselines and excessive visibility amplitudes (mostly RFI) resulted in a 6.75~MHz effective bandwidth. This was followed by amplitude, phase and bandpass calibration on 3C~286. We adopted a Stokes I flux density of 31.01~Jy on the Perley--Taylor 1999.2 scale\footnote{Defined in the VLA Calibrator Manual; \url{http://www.vla.nrao.edu/astro/calib/manual/baars.html}}, which is derived from the \citet{baars1977} scale for 3C~295 (the flux scale is discussed in more detail in Section~\ref{sec:bootes_astro_flux}). To reduce the data volume, each 4~consecutive frequency channels were combined to form 27 channels of 0.25~MHz each. After initial imaging of the target field (see Table~\ref{tab:bootes_imaging}), the calibration on the target field data was improved by three rounds of phase-only self-calibration and one round of amplitude \& phase self-calibration. To remove excessive visibilities at a lower level, the obtained source model was temporarily subtracted from the visibilities, after which the residual visibilities were manually and semi-automatically flagged per baseline. After re-imaging, the noise (RMS of the image background) in the inner half of the (uncorrected) primary beam area, was approximately 1.4~\mjybeam{}. Near the brightest three sources (with apparent flux densities larger than 1~Jy), the local noise was measured to be $\gtrsim 2.2$~\mjybeam{}.

\ctable[botcap,center,
caption = {Overview of wide-field imaging parameters (upper panel) and SPAM parameters (lower panel).},
label = tab:bootes_imaging
]{l c}{
\tnote[a]{We map more than twice the HPBW diameter to allow for deconvolution of nearby bright sources.}
\tnote[b]{\citet{briggs1995}}
\tnote[c]{\citet{perley1989a,cornwell1992}}
\tnote[d]{\citet{conway1990}}
\tnote[e]{\citet{schwab1984,cotton1999,cornwell1999}}
\tnote[f]{\citet{intema2009a}}
\tnote[g]{Specified for the final (second) calibration cycle only.}
\tnote[h]{Power-law slope of the assumed phase structure function.}
\tnote[i]{\citet{condon1994,condon1998}}
}{\FL
Field diameter\tmark[a] & $6.8\degs$ \NN
Pixel size & $4\arcsec$ \NN
Weighting\tmark[b] & robust 0.5 \NN
Wide-field imaging\tmark[c]\tmark[d] & polyhedron (facet-based), \NN
 & multi-frequency synthesis \NN
Number of facets & 199 \NN
Facet diameter & $0.57\degs$ \NN
Facet separation & $0.47\degs$ \NN
Deconvolution\tmark[e] & Cotton-Schwab CLEAN \NN
CLEAN box threshold & $5\,\sigma$ \NN
CLEAN depth & $2\,\sigma$ \NN
Restoring beam & $26\arcsec \times 22\arcsec$ (PA $78\degs$) \LL
SPAM\tmark[f] calibration cycles & 2 \NN
Peeled sources\tmark[g] & 24 \NN
Layer heights (and weights) & 100~km (0.25) \NN
                            & 200~km (0.50) \NN
                            & 400~km (0.25) \NN
Parameter $\gamma$ (\tmark[h]) & 5 / 3 \NN
Model parameters & 20 \NN
Model fit phase RMS\tmark[g] & $19.2\degs \pm 3.0\degs$ \NN
Peeling corrections applied & yes \NN
Reference catalog\tmark[i] & NVSS
\LL
}

Despite the good overall quality of the self-calibrated image from the traditional calibration, there were significant artifacts present in the background near bright sources, limiting the local dynamic range to a few hundred. To suppress artifacts related to direction-dependent calibration errors, we applied the SPAM algorithm on the self-calibrated(!) visibility data. The used SPAM parameters are reported in Table~\ref{tab:bootes_imaging} \citep[for details see][]{intema2009a}. This unconventional combination of self-calibration and SPAM is used to bypass the problem of antenna-based phase discontinuities, as observed in the calibration solutions for several antennas. Applying SPAM after self-calibration can be considered as a perturbation to the overall gradient correction found by self-calibration. The resulting noise in the inner half of the (uncorrected) primary beam area is on average 1.0~\mjybeam{}. The local noise near the brightest three sources is reduced to $\lesssim 2.0$~\mjybeam{}.

For an observation of $\sim 10$~hours, the theoretical thermal noise level\footnote{Derived using the GMRT User Manual; \url{http://gmrt.ncra.tifr.res.in/gmrt_hpage/Users/doc/manual/UsersManual/index.html}} for the GMRT at 153~MHz is estimated to be $\sim 0.2 - 0.3$~\mjybeam{}. For our observation, the measured noise of $\sim 1.0$~\mjybeam{} in the central part of the field is a factor of 3 to 5 larger. In apparently source-free parts of the outer imaged area this drops to 0.75~\mjybeam{}, indicating that residual sidelobe structure has a noticable contribution to the central noise. Although significantly improved as compared to the self-calibrated case, the residual sidelobe structure near the three bright sources, all located near the half-power radius of the primary beam area, limits the local dynamic range to $\sim 600$. We expect that the residual sidelobes are caused by direction-dependent visibility amplitude errors due to pointing errors and non-circular primary beam patterns \citep[e.g.,][]{bhatnagar2008}.

Another suspect for adding to the overall noise is residual RFI (see also Section~\ref{sec:bootes_astro_flux}). Especially on the most affected, short baselines, it can be difficult to distinguish between persistent broad-band RFI and true sky signal. In our approach we have attempted to save as many short baselines as possible, to preserve the sensitivity to large-scale emission. This could explain why our central noise level of $\sim 1.0$~\mjybeam{} is slightly higher than the quoted 0.7~\mjybeam{} found by \citet{sirothia2009,ishwara2010} for similar deep fields. In general, we find that a straightforward comparison of noise levels is non-trivial, as it depends on the flagging \& data selection strategy, the visibility weighting scheme used during imaging, and on how and where in the image the noise is measured.

\section{Catalog Construction}
\label{sec:bootes_catalog}

\begin{figure}[!tp]
\begin{center}
\resizebox{\hsize}{!}{\includegraphics[angle=0]{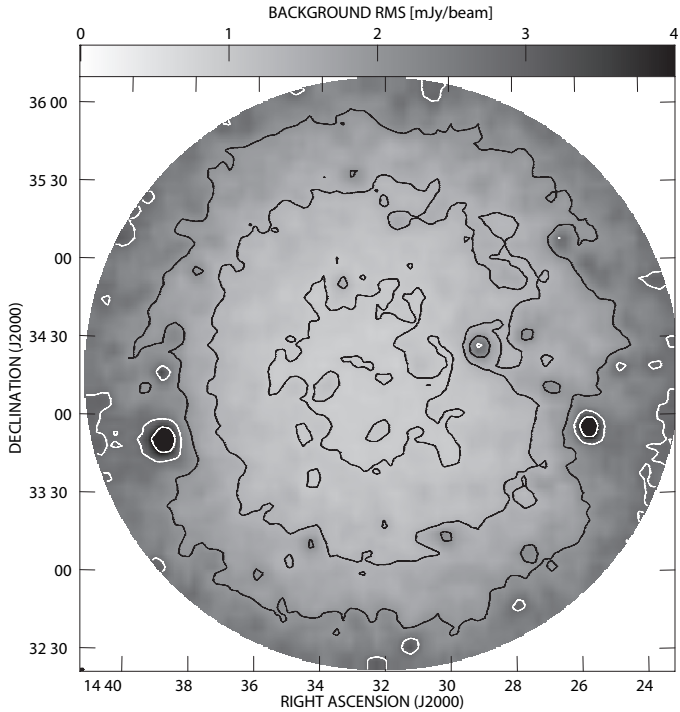}}
\caption{The grayscale map represents the local noise as measured in the (primary beam corrected) \bootes{} field image. The overplotted contours mark lines of equal RMS at $[ 1.0, 1.4, 2.0, 2.8, 4.0 ]$~\mjybeam{} (white above 2.0~\mjybeam{}). The local enhancements in RMS coincide with the positions of the brightest sources.}
\label{fig:bootes_rms_map}
\end{center}
\end{figure}

In this section we describe the construction of the source catalog that was extracted from the 153~MHz image of the \bootes{} field. The image after SPAM calibration was scaled down by 4~percent to incorporate a correction to the flux calibrator scale (see Section~\ref{sec:bootes_astro_flux}). Next, the image was corrected for primary beam attenuation with a circular beam model\footnote{See the GMRT User Manual} out to a 1.9~degree radius (out to 0.3 times the central beam response, slightly beyond the half-power radius), yielding a survey area of 11.3~square degrees. Figure~\ref{fig:bootes_rms_map} shows a map of the local noise of the circular survey area, which has a central area with a noise level of $\sim 1.0$~\mjybeam{}, a global increase of the local noise to 2.0--2.5~\mjybeam{} when moving towards the field edge, and several small areas around bright sources where the local noise is approximately twice the surrounding noise level. The overall noise was found to be 1.7~\mjybeam{}.

Although we used multi-frequency synthesis, the 0.25~MHz width of individual frequency channels causes bandwidth smearing during imaging. Applying the standard formula \citep[e.g.,][]{thompson1999} to our case, the radial broadening of sources at the edge of the field (at 1.9~degree from the pointing center) is estimated to be $\sim 11\arcsec$, which is half the minor axis of the restoring beam. Similarly, the visibility time resolution of 16.8~sec causes time-average smearing in the order of $\sim 8\arcsec$ at the field's edge. This may appear problematic, but the majority of sources are detected within the inner part of the primary beam (both effects scale linearly with radial distance from the pointing center). The total flux density of smeared sources is conserved, but the peak flux drops. The effect of smearing on source extraction is discussed in Section~\ref{sec:bootes_compl}.

\subsection{Source Extraction}
\label{sec:bootes_bdsm}

We used the BDSM software package \citep{mohan2008} to extract sources from our image. With our settings, BDSM estimates the local background noise level $\sigma^{}_\mathrm{L}$ over the map area, searches for pixels $> 5 \, \sigma^{}_\mathrm{L}$, expands the $5 \, \sigma^{}_\mathrm{L}$ detections into islands by searching for adjacent pixels $> 3 \, \sigma^{}_\mathrm{L}$, rejecting islands smaller than 4~pixels. BDSM fits the emission in each island with one or multiple gaussians, combines significantly overlapping gaussians into sources, and determines the flux densities, shapes and positions of sources \citep[including error estimates, following][]{condon1997}. This approach has led to very few false source detections (see Section~\ref{sec:bootes_compl}).

BDSM detected 644~islands, for which the 935~fitted gaussians were grouped into 696~distinct sources. Of these, 499~sources were fitted with a single gaussian. Visual inspection of the image, complemented with a comparison against a very deep 1.4~GHz map (see Section~\ref{sec:bootes_astro_flux}), resulted in the removal of 16~source detections in the near vicinity of six of the seven brightest sources. We also removed 4~sources that extended beyond the edge of the image. To facilitate total flux density measurements at the high flux end, we combined multiple source detections in single islands, and manually combined 50~additional source detections that were assigned to different islands but appeared to be associated in either the 153~MHz image or the deep 1.4~GHz map. The combined flux density is the sum of the individual components, while the combined position is a flux-weighted average (centroid). Error estimates of the positions and flux densities of the components are propagated into error estimates of the combined flux density and position. The final catalog consists of 598~sources.

Table~\ref{tab:bootes_catalog} contains a sample of the full source catalog available in the CDS\footnote{\url{http://cdsweb.u-strasbg.fr/}} database. The source entries include the corrections discussed in the remainder of Section~\ref{sec:bootes_catalog}. The columns are: (1) source identifier, (2,3) centroid position right ascension and declination, (4,5) centroid position uncertainty along the right ascension and declination directions, (6,7) integrated source flux and uncertainty, (8) local background RMS noise level, and (9) number of gaussians fitted to the source.

\ctable[botcap,center,star,
caption={Sample of the full source catalog of GMRT 153~MHz sources. The columns are described in the text.},
label=tab:bootes_catalog
]{c c c c c c c c c c}{
}{\FL
ID & RA & DEC & $\Delta$RA & $\Delta$DEC & $S_\mathrm{153}$ & $\Delta{}S_\mathrm{153}$ & Noise & $N_\mathrm{gauss}$ \NN
{} & [$\degs$] & [$\degs$] & [$\arcsec$] & [$\arcsec$] & [mJy] & [mJy] & [\mjybeam{}] & {} \NN
(1) & (2) & (3) & (4) & (5) & (6) & (7) & (8) & (9)\ML
J142632+350815 & 216.63367 & 35.13754 &  5.2 &  4.3 &  649.7 & 65.1 & 2.8 & 1 \NN
J143810+340459 & 219.54576 & 34.08315 &  3.9 &  2.8 &  641.2 & 64.2 & 2.0 & 1 \NN
J143850+330225 & 219.71068 & 33.04036 & 10.0 &  8.8 &  577.8 & 58.2 & 2.9 & 2 \NN
J143134+351505 & 217.89466 & 35.25165 &  5.2 &  3.5 &  567.8 & 56.9 & 1.6 & 2 \NN
J142430+343630 & 216.12544 & 34.60836 & 71.1 & 60.3 &  543.0 & 55.2 & 2.5 & 7 \NN
J142445+341816 & 216.19020 & 34.30447 &  7.2 &  5.3 &  507.1 & 50.9 & 2.9 & 1 \NN
J143055+341933 & 217.73287 & 34.32597 &  3.9 &  3.0 &  505.0 & 50.6 & 1.2 & 2 \NN
J143256+353339 & 218.23353 & 35.56104 &  4.7 &  4.0 &  503.8 & 50.5 & 2.0 & 1 \NN
J143930+344628 & 219.87580 & 34.77470 &  9.2 &  7.3 &  489.8 & 49.4 & 2.5 & 2 \NN
J143128+355252 & 217.86699 & 35.88132 & 11.6 & 11.6 &  459.4 & 46.5 & 2.3 & 3 \LL
}

\subsection{Completeness and Contamination}
\label{sec:bootes_compl}

\begin{figure*}
\begin{center}
\resizebox{\hsize}{!}{
\includegraphics[width=0.1\hsize,angle=0]{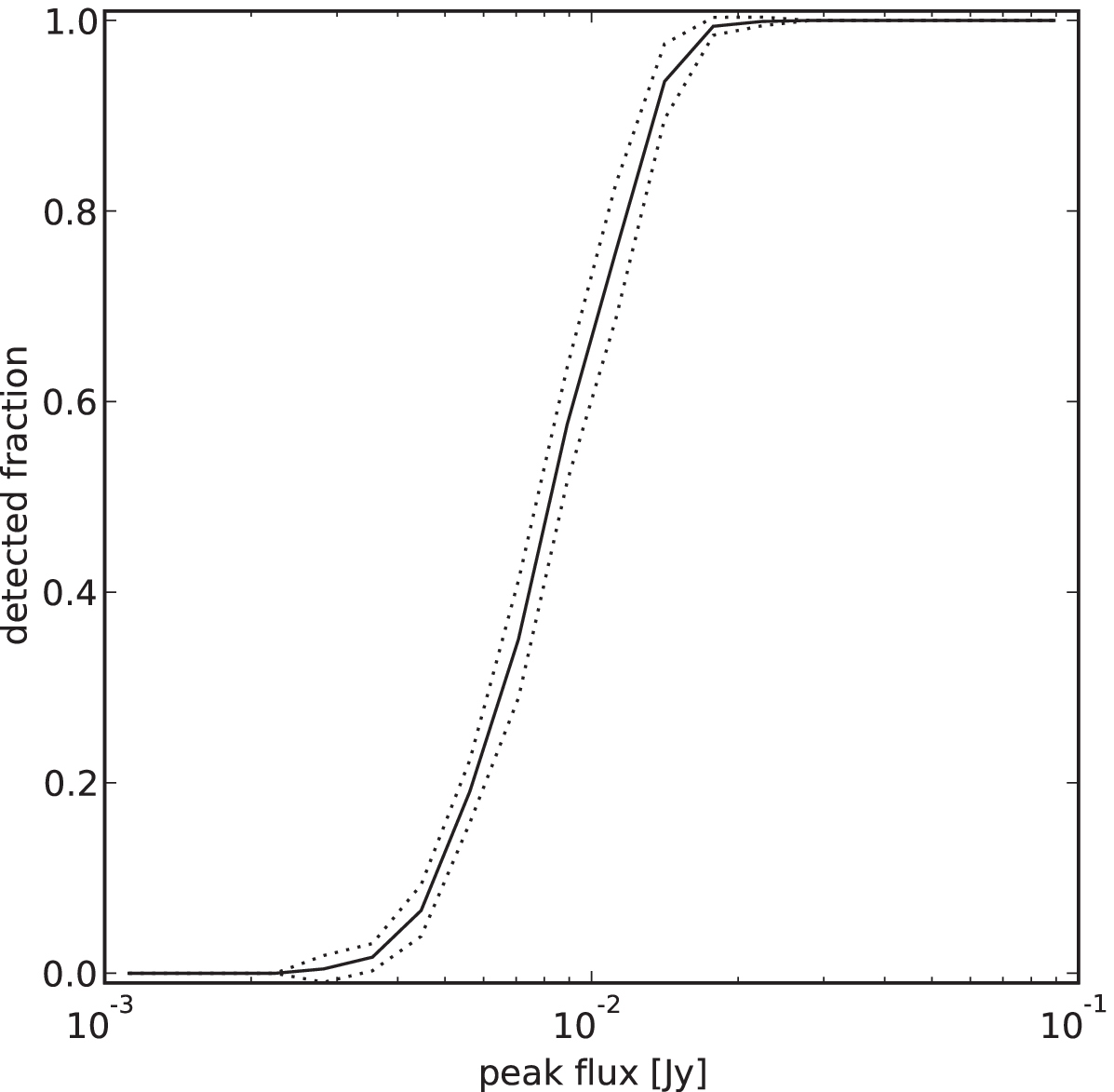}
\includegraphics[width=0.1\hsize,angle=0]{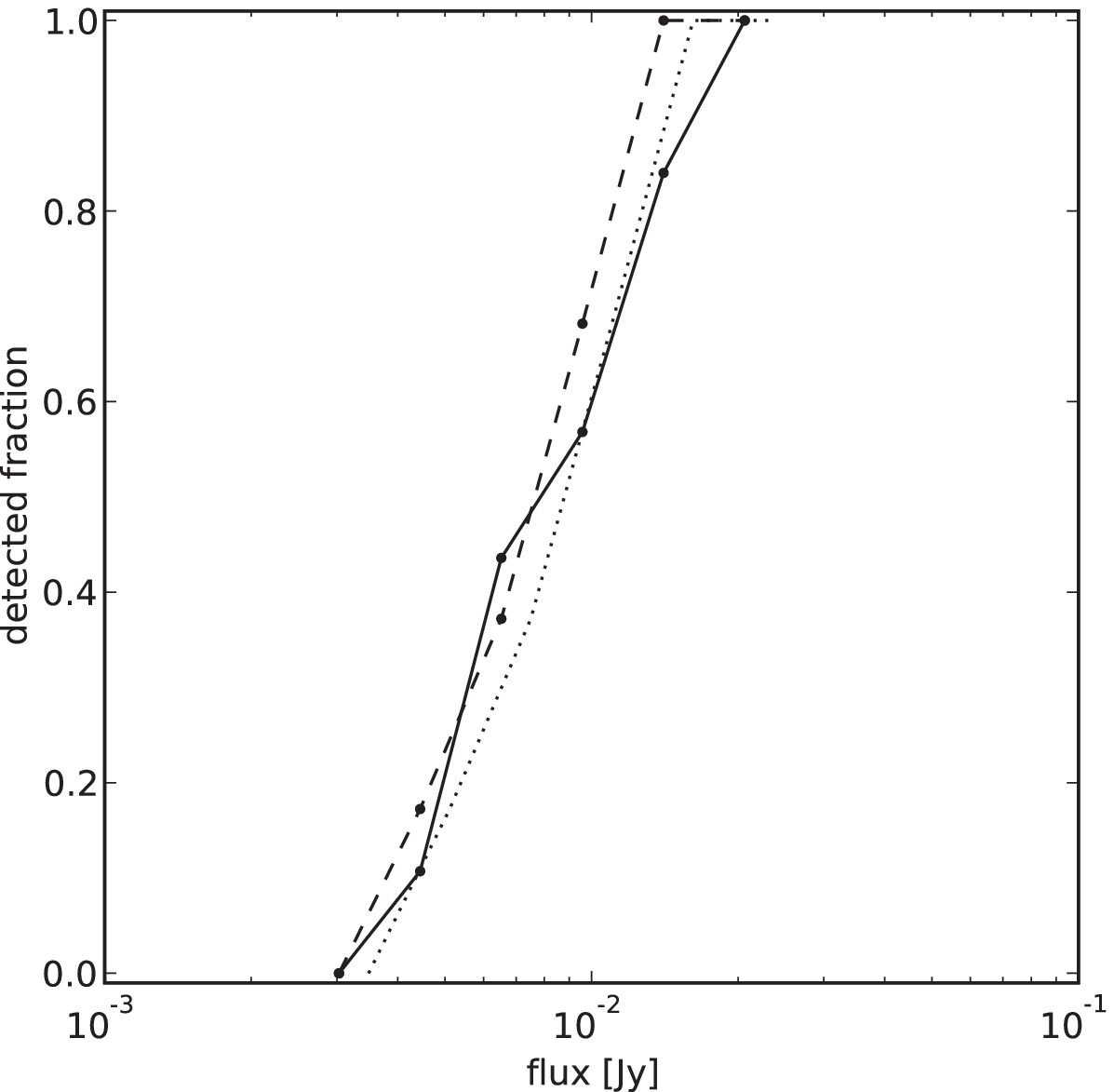}
\includegraphics[width=0.1\hsize,angle=0]{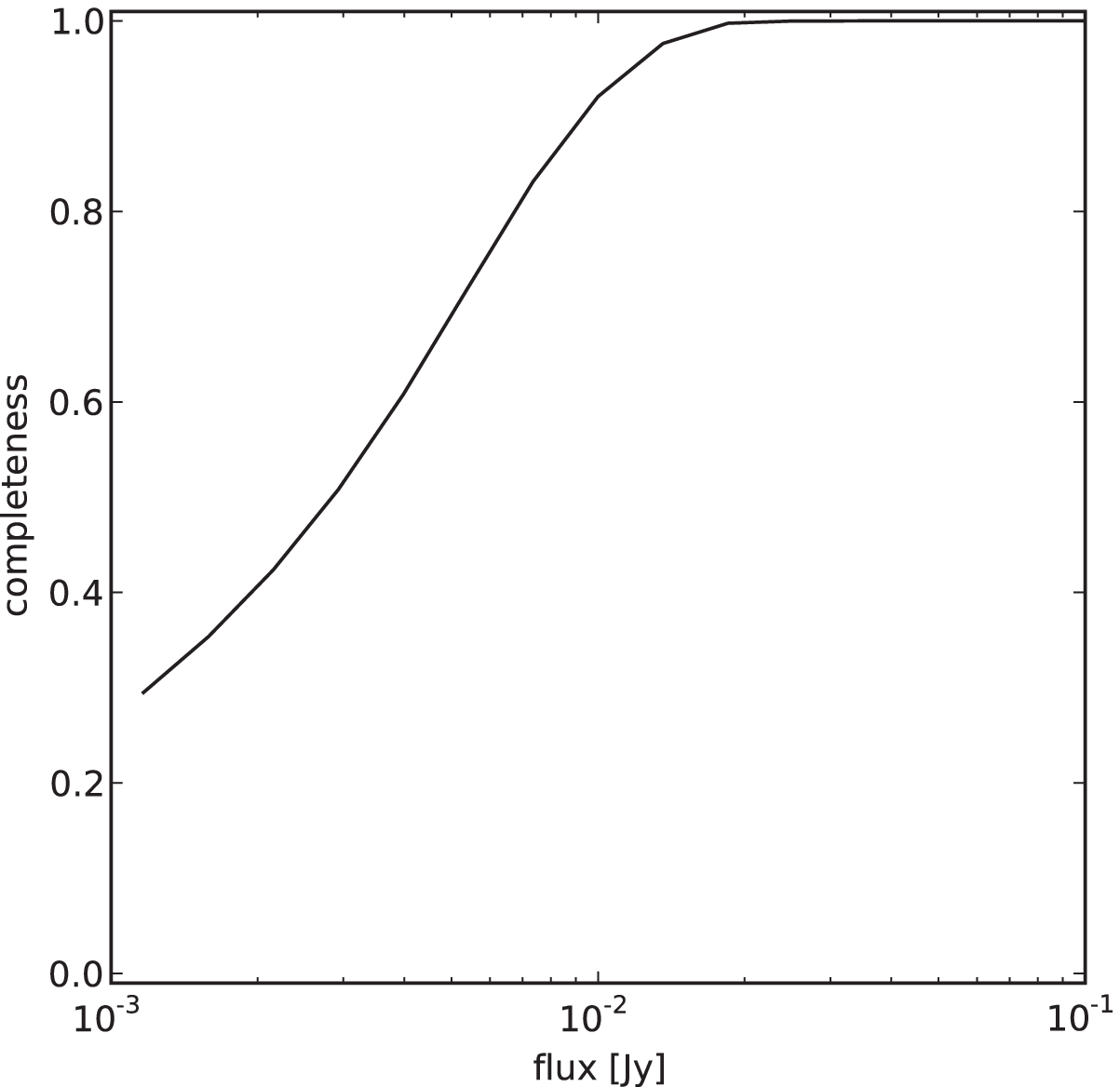}}
\caption{Detection fractions and completeness as a function of flux. \textit{Left}: Result from 20 Monte-Carlo simulations, in which 1000~point sources with varying fluxes were inserted into the residual image, followed by a source extraction. The horizontal axis denotes the input (peak) flux. The solid line is the average detection fraction, and the dotted lines denote the $1 \,\sigma$ uncertainty. \textit{Middle}: Result from scaling the peak flux and total flux density of a subset of high S/N sources with varying sizes down to the detection limit, and applying the $5 \, \sigma^{}_\mathrm{L}$ detection criteria. The dashed line is the detection fraction as a function of peak flux, the solid line is the detection fraction as a function of total flux density, and the dotted line is the dashed line shifted in flux by 15~percent to approximately match the solid line. \textit{Right}: Estimated completeness of the extracted source catalog as a function of (total) flux limit.}
\label{fig:bootes_detection}
\end{center}
\end{figure*}

We estimated the completeness of the 153~MHz catalog by performing Monte-Carlo simulations. The source extraction process generated a residual image from which all detected source flux was subtracted. For our simulation, we inserted 1000~artificial point sources into the residual image, and used the same mechanism as described in Section~\ref{sec:bootes_bdsm} to extract them. The artificial source positions were selected randomly, but never within $50\arcsec$ of another source, a blanked region (near the image edge) or a $>10$~mJy residual. The source peak fluxes $S$ were chosen randomly within the range 3~mJy to 3~Jy, while obeying the source count statistic $dN / dS \propto S^{-1.5}_{}$, which produced statistically sufficient detection counts in all logarithmic flux bins. The simulation was repeated 20~times to improve the accuracy and to derive error estimates. The detection fraction as a function of peak flux is plotted in Figure~\ref{fig:bootes_detection}.

In this approach we have ignored several effects that may influence the detectability of sources, such as (i) the intrinsic size of sources, (ii) bandwidth- and time-average smearing, (iii) calibration errors, and (iv) imaging and deconvolution. The source detection algorithm uses a peak detection threshold. Sources that are resolved, smeared due to bandwidth- and time-averaging, or defocussed due to calibration errors will therefore have a decreased probability of detection. We have not attempted to model for the angular size distribution of sources at this frequency. However, previous observations show that the major fraction of low-frequency sources are unresolved at $\sim 25\arcsec$ resolution \citep[e.g.,][]{cohen2004,george2009}. In our catalog, more than 90~percent of the sources appear to have simple, near-gaussian morphologies. Assuming the angular size distribution of sources changes slowly with source flux density, and assuming that smearing and defocussing affects sources of varying brightness in a statistically equal way, we can estimate the resolution bias from the catalog itself. For this purpose, we select a subset of 214~high S/N sources with peak fluxes between 12 and $20\,\sigma^{}_\mathrm{L}$ and simple morphologies. The flux densities of these sources were scaled down by a factor of 4 to create an artificial population of sources near the detection threshold. After applying the $5 \, \sigma^{}_\mathrm{L}$~detection criterium, we determined the detection fractions, both as a function of peak flux and total (integrated) flux density (Figure~\ref{fig:bootes_detection}). Although this approach suffers from low number statistics, the general trend of both detection fraction functions is similar but shifted upwards in flux by $\sim 15$~percent. We therefore assume that the total flux detection fraction is approximated by the peak flux detection fraction derived from the Monte-Carlo simulations, shifted upwards in flux by 15~percent.

To estimate the completeness of the catalog at various flux limits, the total flux detection fraction must be multiplied by the (normalized) true source flux distribution, and integrated from the flux limit upwards. The Euclidean-normalized differential source counts in Section~\ref{sec:bootes_dsc} approximately follow a power-law over the larger part of the flux range of our sample, with a slope of 0.91. The source flux distribution over the same flux range can therefore be approximated by a power-law with slope of $0.91 - 5/2 = -1.59$. As it involves very few counts, the deviation of the power-law from the true source counts at the high flux end has little effect on our estimates, and is therefore ignored. Figure~\ref{fig:bootes_detection} shows the resulting completeness estimate as a function of total flux. From this plot, the estimated completeness is $\sim 70$~percent at 5.1~mJy, $\sim 92$~percent at 10~mJy and $> 99$~percent at 20~mJy.

A known effect that arises from the deconvolution process is CLEAN bias \citep[e.g.,][]{condon1994,condon1998,becker1995}. This is a systematic negative offset in the recovered flux densities after deconvolution, probably the result of false CLEANing of sidelobe peaks in the dirty beam pattern. One can estimate the CLEAN bias by injecting artificial sources into the visibility data and compare the recovered flux densities after imaging \& deconvolution with the injected flux densities. We have not attempted this approach, but instead taken precautions to minimize the CLEAN bias effect. In our case, the dirty beam is well-behaved due to a relatively uniform UV-coverage from two extended observing runs, in combination with multi-frequency synthesis and a robust weighting parameter of 0.5 (slightly towards natural weighting). CLEAN bias is further suppressed through the use of tight CLEAN boxes in the imaging \& deconvolution process.

For an estimate of the contamination of the catalog with fake detections, we compare the GMRT 153~MHz image and extracted source catalog against the results from the deep WSRT 1.4~GHz survey of the \bootes{} field by \citet{devries2002}. The 153~MHz and 1.4~GHz observations are well matched in terms of survey area and resolution ($\sim 7$~square degrees and $13\arcsec \times 21\arcsec$ for WSRT, respectively). The typical noise over the WSRT survey area is 28~$\mu$\jybeam. For a spectral index of $-0.8$, the 1.4~GHz observations are $\sim 10$~times more sensitive. We restrict our comparison to a 1.4~degree radius circular area to avoid the noisy edge of the deep 1.4~GHz survey. For all of the 399~sources detected at 153~MHz we find a counterpart in the 1.4~GHz map (383~sources were automatically matched within a $25\arcsec$ search radius, while the remaining fraction of sources with complex morphology were confirmed manually). We could not match the full GMRT area, but considering that our source extraction is based on local noise and that the false detections only appeared to occur near a few bright sources, we estimate that the contamination of our complete catalog over the full survey area is $< 1$~percent.

\subsection{Astrometric and Flux Uncertainty}
\label{sec:bootes_astro_flux}

For an estimate of the astrometric uncertainty, we compare the source positions in the GMRT 153~MHz catalog against catalog source positions from the deep WSRT 1.4~GHz map of \citet{devries2002}. For our position comparison, we only use sources whose flux profile is accurately described by a single gaussian, and whose peak flux $S_p$ is at least $10\,\sigma^{}_\mathrm{L}$. This bypasses most of the position errors that arise from low signal-to-noise (S/N, or $S_p / \sigma^{}_\mathrm{L}$), different grouping of gaussians and spectral variations across sources. Using a search radius of $10\arcsec$, we cross-match 126~sources in both catalogs. From this, we measure a small mean position offset in RA and DEC of $(\Delta\alpha,\Delta\delta) = (0.11\arcsec,0.09\arcsec)$. We correct the catalog positions for this small offset. The estimated RMS scatter around this offset is $\sigma^{}_{\alpha,\delta} = 1.32\arcsec$. The S/N-independent part of the positional uncertainty of the 1.4~GHz sources is $0.44\arcsec$ \citep{devries2002}, therefore we derive an absolute astrometric uncertainty for the 153~MHz sources of $1.24\arcsec$. We quadratically add this to the calculated (S/N-dependent) position accuracies in the 153~MHz source catalog (see Section~\ref{sec:bootes_bdsm}).

The uncertainty of the flux scale transferred from the calibrator 3C~286 to the target \bootes{} field is influenced by several factors: (i) the quality of the calibrator observational data, (ii) the accuracy of calibrator source model, and (iii) the difference in observing conditions between the calibrator and target field. Because of the relatively large uncertainty in the flux scale at low frequencies, we discuss in some detail the issues that influence these factors.

The quality of the calibrator data is most noticeably affected by RFI and by ionospheric phase rotations. The repeated observation of 3C~286 every $\sim 45$~minutes during the observing enabled us to monitor these effects over time. The mild fluctuations in the initial (short interval) calibration gain phases at the start of the data reduction showed that the ionosphere was very calm during both observing nights, therefore we exclude the possibility of diffraction or focussing effects \citep[e.g.,][]{jacobson1992a}. Apparent flux loss due to ionospheric phase rotations was prevented by applying the (short interval) gain phase corrections before bandpass- and amplitude calibration (see Section~\ref{sec:bootes_dr}).

RFI was continuously present during both observing sessions. This mainly consisted of persistent RFI over the full band, most noticeably on the shortest (central square and neighbouring arm antenna) baselines, and of more sporadic events on longer baselines during one or more time stamps and/or narrow frequency ranges. The sporadic events were relatively easy to recognize and excise, but for the persistent RFI this is much more difficult due to a lack of contrast between healthy and affected data on a single baseline. Some of the shortest, most affected baselines were removed completely. On longer baselines, persistent RFI from quasi-stationary sources can average out due to fringe tracking \citep{athreya2009}, but does add noise. Large magnitude RFI amplitude errors in the visibilities may result in a suppression of the gain amplitude corrections. Because these effects are hard to quantify, we adopt an ad-hoc 2~percent amplitude error due to RFI.

\begin{figure}[!tp]
\begin{center}
\resizebox{\hsize}{!}{\includegraphics[angle=0]{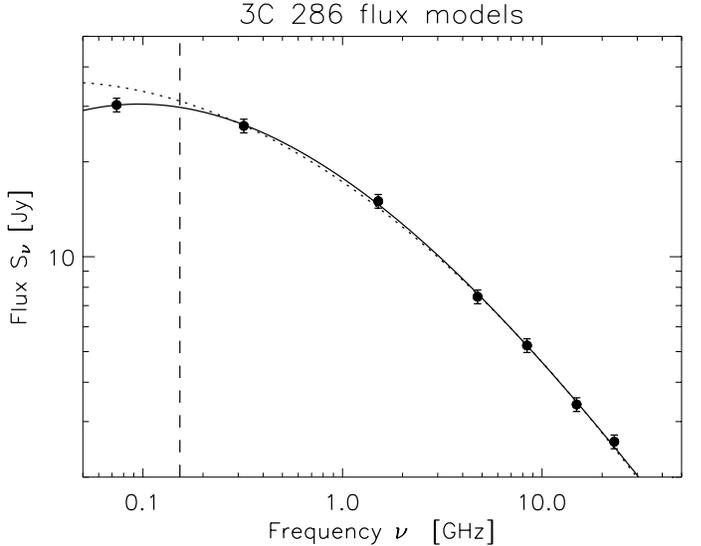}}
\caption{Model of the radio spectrum of 3C~286. The measurements (dots) above 153~MHz (dashed line) were used to fit the original Perley--Taylor 1999.2 model (dotted line), which appears to overestimate the flux density at 153~MHz. By adding one measurement at 74~MHz, we fit an alternative model (solid line) that may give a better prediction at 153~MHz.}
\label{fig:bootes_flux_3c286}
\end{center}
\end{figure}

\ctable[botcap,center,star,
caption = {Flux measurements of 3C~286 from different radio survey catalogs at low frequencies.},
label = tab:bootes_fluxes_3c286
]{c c c c c}{
\tnote[a]{Perley-Taylor}
\tnote[b]{\citet{cohen2007}}
\tnote[c]{\citet{hales1988}}
\tnote[d]{\citet{hales2007} and references therein. Note that we use the peak flux rather than the integrated flux density, as 3C~286 is unresolved on the 7C resolution.}
\tnote[e]{\citet{edge1959,bennett1962}}
\tnote[f]{\citet{pilkington1965,gower1967}}
\tnote[g]{\citet{rengelink1997}. Note that we use the peak flux rather than the integrated flux, as 3C~286 is unresolved on the WENSS resolution.}
}{\FL
Survey & Frequency & Catalog flux & PT\tmark[a] model flux & New model flux \NN
               & [MHz] & [Jy]           & [Jy]  & [Jy]  \ML
VLSS\tmark[b]  &  74   & $30.26\pm3.08$ & 34.67 & 30.26 \NN
6C\tmark[c]    & 151   & $26.31$        & 31.25 & 29.81 \NN
7C\tmark[d]    & 151   & $26.53$        & 31.25 & 29.81 \NN
3C\tmark[e]    & 159   & $30.0\pm7.0$   & 30.94 & 29.65 \NN
4C\tmark[f]    & 178   & $24.0$         & 30.22 & 29.26 \NN
WENSS\tmark[g] & 325   & $27.12\pm1.63$ & 25.96 & 26.11 \LL
}

Because 3C~286 is unresolved ($\lesssim 2.5\arcsec$) within a $20\arcsec$ to $25\arcsec$ beam, we used a point source for the calibrator model with a flux density of 31.01~Jy (Section~\ref{sec:bootes_dr}). The utilized Perley--Taylor flux density at 153~MHz is an extrapolatation of VLA flux density measurements at 330~MHz and higher, using a fourth order polynomial in log-log space. In Table~\ref{tab:bootes_fluxes_3c286} a comparison is presented between flux density measurements of 3C~286 in various sky surveys at low frequencies and the predicted flux densities from the Perley--Taylor model. Although there is a large variation in the flux differences at the different frequencies, there appears to be a overestimation by the Perley--Taylor model below 200~MHz due to a spectral turnover of 3C~286 below $\sim 300$~MHz. For this reason, we re-fitted the polynomial to the original data points plus the additional 74~MHz VLSS measurement, assuming 5~percent errors for all data points (Figure~\ref{fig:bootes_flux_3c286}). Our new model is given by:
\begin{eqnarray}
\log^{}_{10}(S^{}_{\nu}) && = 1.24922 - 0.434710 \log^{}_{10}(\nu) \nonumber \\
- && 0.174786 ( \log^{}_{10}(\nu))^{2}_{} + 0.0251542 ( \log^{}_{10}(\nu) )^{3}_{},
\label{eq:flux_3c286}
\end{eqnarray}
with $\nu$ the frequency in GHz and $S^{}_{\nu}$ the flux density in Jy. This parametrization results in a slightly better fit to the flux densities from the various surveys (Table~\ref{tab:bootes_fluxes_3c286}), although the large scatter remains. For the center of the GMRT band at 153.1~MHz, the new model predicts a flux density of 29.77~Jy. We have adopted this new model by scaling the image by the ratio $29.77 / 31.01$ before primary beam correction (see start of Section~\ref{sec:bootes_catalog}). The accuracy of this flux scale is closely related to the accuracy of the 74~MHz data point, for which a 10~percent flux error is given. Assuming the error in the 330~MHz is much smaller, and 153~MHz is roughly half way from 74 to 330~MHz in logarithmic frequency, we set an upper limit of 5~percent error on the adopted flux scale of 3C~286 at 153~MHz.

The presence of other sources in the 3.1~degree FoV around 3C~286 further complicates the flux calibration. For example, the total apparent flux density in the \bootes{} field that was extracted through CLEAN deconvolution is $\sim 46$~Jy, which may be typical lower limit for any blind field. If this flux is distributed over many sources that are individually much fainter than the calibrator, then the net effect of this additional flux is only noticeable on a small subset of the shortest baselines, while calibration utilizes all baselines. Inspection of the 3C~286 field at 74~MHz \citep[VLSS;][]{cohen2007} and 325~MHz \citep[WENSS;][]{rengelink1997} does identify two relatively bright sources within 0.7~degrees of 3C~286 with estimated apparent GMRT 153~MHz flux densities of 5.3 and 2.7~Jy, respectively. These sources will cause a modulation of the visibility amplitudes across the UV-plane. We performed a simple simulation, in which we replaced the measured 3C~286 visibilities with noise-less model visibilities of three point sources, being 3C~286 and the two nearby sources, and calibrated these visibilities against a single point source model of 3C~286, using the same settings as in the original data reduction. We found that the combined gain amplitude for all antennas and all time intervals was $1.000 \pm 0.004$. For individual 10~minute time blocks, the largest deviation from one was 1.00~percent, which indicates the magnitude of the possible error when using a single 10~minute calibrator observation on 3C~286. The small deviations per time block are transferred to the target field data, but these are suppressed by amplitude self-calibration against the target field source model. We set an upper limit of 1~percent due to the presence of other sources in the FoV of the calibrator.

While transferring the flux scale from calibrator to target field, the derived gain amplitudes need to be corrected for differences in sky temperature due to galactic diffuse radio emission \citep[e.g.,][]{tasse2007}, which is detected by individual array antennas but not by the interferometer. The GMRT does not implement a sky temperature measurement, therefore we need to rely on an external source of information. From the \citet{haslam1982} all-sky map, we find that the mean off-source sky temperatures at 408~MHz as measured in the 3C~286 and \bootes{} field are both approximately $20 \pm 1$~degree. Applying the formulae given by \citet{tasse2007} for the GMRT at 153~MHz\footnote{We adopted the GMRT system parameters from \url{http://www.gmrt.ncra.tifr.res.in/gmrt_hpage/Users/doc/manual/UsersManual/node13.html}}, we estimate a gain inaccuracy of $\sim 2$~percent at most. 

Another effect that may influence the gain amplitude transfer between calibrator and target field is an elevation-dependent gain error. This is a combination of effects such as structural deformation of the antenna, atmospheric refraction and changes in system temperature from ground radiation. According to \citet{chandra2004}, the effect on amplitude is rather small, if not negligible, for GMRT frequencies of 610~MHz and below. Furthermore, the relatively short angular distance of 13.4~degrees between 3C~286 and the target field center causes the differential elevation error to be limited. Elevation dependent \textit{phase} errors are not relevant, because we don't rely on the calibrator to restore the astrometry. For our observation, we assume that elevation-dependent effects can be ignored.

To incorporate the effects discussed above plus some margin, we quadratically add a relative flux error of 10~percent to the flux density measurement errors in the source catalog.

\section{Analysis}
\label{sec:bootes_analysis}

The 598~radio sources in our 153~MHz catalog form a statistically significant set, ranging in flux density from 5.1~mJy to 3.9~Jy. Because of the large survey area, cosmic variance is expected to be small. Radio images at 153~MHz of a selection of extended sources are presented in Appendix~\ref{sec:bootes_images}. In this section, we discuss two characteristics of the survey: number counts and spectral indices. Further analyses will be done in a subsequent article. When properly corrected for incompleteness, the number counts are an objective measure of the 153~MHz source population, that can be compared against models and other surveys. For individual sources, the spectral index can help to classify sources, and identify rare sources such as HzRGs and cluster halos \& relics.

\subsection{Differential Source Counts}
\label{sec:bootes_dsc}

\begin{figure}[!tp]
\begin{center}
\resizebox{\hsize}{!}{\includegraphics[angle=0]{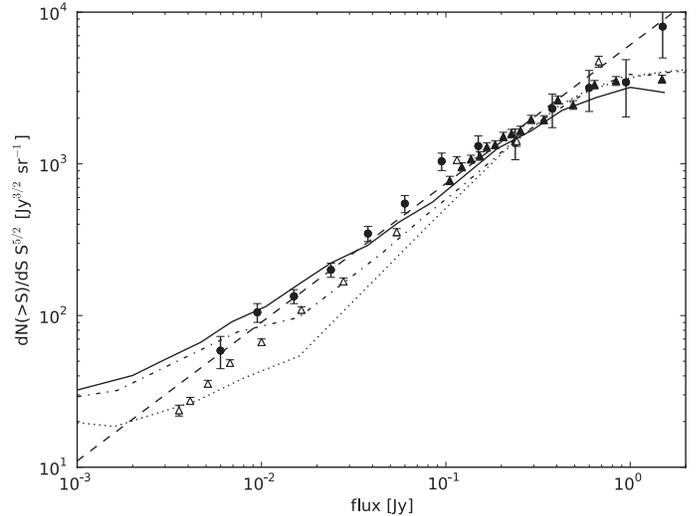}}
\caption{Euclidean-normalized differential source counts for the full 153~MHz catalog of 598~sources, distributed over 14~logarithmic flux bins ranging from 4.75~mJy to 3~Jy (black dots; we have omitted the highest flux bin, containing just one count). Also plotted are 151~MHz source counts by \citet{mcgilcrist1990}, part of the 7C catalog (black triangles), 153~MHz source counts by \citet[][white triangles]{ishwara2010}, model 151~MHz source counts by \citet{jackson2005} for a $\Lambda$CDM cosmology (dotted line) and an $\Omega_\mathrm{m}=1$ cosmology (dash-dotted line), a 151~MHz source count model by \citet[][solid line]{wilman2008}, and the power-law slope of 0.91 (dashed line) that fits our data points best.}
\label{fig:bootes_dif_source_counts}
\end{center}
\end{figure}

We derived the Euclidean-normalized differential source counts from the catalog. Because the source extraction criteria vary over the survey area, we used Figure~\ref{fig:bootes_detection} to correct for the missed fraction per flux bin. This mainly affects the lowest two bins. Furthermore, the combined effect of random peaks in the background noise and a peak detection criteria causes a selection bias for positively enhanced weak sources (Eddington bias). In general, noise can scatter sources into other flux bins, most noticeably near the detection limit. Our attempts to correct for this effect through Monte-Carlo simulations were numerically unstable due to the low number counts in the lowest flux bin. The effect on the higher bins was minimal, therefore we omit the Eddington bias corrections. The Euclidean normalized differential source counts are tabulated in Table~\ref{tab:bootes_dif_counts}. Figure~\ref{fig:bootes_dif_source_counts} shows our source counts in comparison with the (little) other observational data that is available for this frequency at comparable sensitivity, as well as source counts derived from two models.

\ctable[botcap,center,
caption = {Euclidean-normalized differential source counts (including poissonian error estimates) for the full 153~MHz catalog of 598~sources, distributed over 14~logarithmic flux bins ranging from 4.75~mJy to 3~Jy.},
label = tab:bootes_dif_counts
]{c c c}{}{\FL
Flux bin center & Raw counts & Normalized counts \NN
{[Jy]} &  & {[Jy$^{3/2}_{}$~sr$^{-1}_{}$]} \ML
0.0060  & 30  & $  58.70 \pm   14.12$ \NN
0.0095  & 89  & $ 105.05 \pm   14.92$ \NN
0.0150  & 102 & $ 133.90 \pm   14.18$ \NN
0.0238  & 87  & $ 199.81 \pm   21.45$ \NN
0.0378  & 76  & $ 347.09 \pm   39.81$ \NN
0.0599  & 60  & $ 546.74 \pm   70.58$ \NN
0.0945  & 57  & $1036.35 \pm  137.27$ \NN
0.1504  & 36  & $1305.98 \pm  217.66$ \NN
0.2383  & 19  & $1375.27 \pm  315.51$ \NN
0.3777  & 16  & $2310.75 \pm  577.69$ \NN
0.5986  & 11  & $3169.75 \pm  955.72$ \NN
0.9487  & 6   & $3449.72 \pm 1408.34$ \NN
1.5036  & 7   & $8030.29 \pm 3035.16$ \NN
2.3830  & 1   & $2288.93 \pm 2288.93$ \LL
}

At the high flux end ($0.1-1$~Jy), there is a good agreement between our source counts and those derived from part of the 7C survey at 151~MHz \citep{mcgilcrist1990}. \citet{ishwara2010} present source counts of 765~sources in a single-field, deep GMRT 153~MHz observation similar to ours, but with over twice the bandwidth (16~MHz). Their source counts go down to $\sim 2.5$~mJy. While their source counts are roughly equal to ours at the high flux end, they are increasingly lower towards lower flux levels. Their source counts are well approximated by a single power-law slope of 1.01, which is steeper than the value of 0.91 from our data over the range 40--400~mJy. This suggests a non-linear flux scale difference. We can offer little explanation on this apparent discrepancy. Within 2--3 times the ($1\sigma$) error bars, the two source count results may still be considered consistent.

\citet{george2008} have determined source counts from a shallower GMRT 153~MHz field (central RMS 3.1~\mjybeam{}) centered around $\epsilon$~Eridanus. From binning 113~source fluxes over a range of 20~mJy--2~Jy they derive a single power-law slope of 0.72 (not plotted). The most likely explanation for this much flatter slope is the large uncertainties in their counts due to low-number statistics.

\citet{jackson2005} constructed a 151~MHz source count model, based on an extrapolation of source counts $> 0.2$~Jy from the 3CRR and 6C catalogs \citep{laing1983,hales1988}. Two cosmological scenarios are considered, namely an $\Omega_\mathrm{m}=1$ cosmology and a $\Lambda$CDM cosmology, the latter being today's generally accepted cosmology \citep[e.g.,][]{komatsu2011}. In this model, \frii{} radio sources dominate the counts above $\sim 50$ and $\sim 20$~mJy, for the two cosmologies respectively. Below, \fri{} sources are the most dominant population down to below our detection threshold, which causes a flattening of the counts below $\sim 20$~mJy in both cosmologies. 

Figure~\ref{fig:bootes_dif_source_counts} shows that the source count models for both scenarios roughly match our observed source counts near $500$~mJy, but increasingly underestimate the counts towards lower flux densities. Between 20 and 200~mJy, the approximately constant model slope for both scenarios is 0.98 and 1.19, respectively, steeper than the value of 0.91 derived from our data. The observed source counts shows no clear evidence of flattening towards lower flux levels.

\citet{wilman2008} have generated model source counts at 151~MHz from a semi-empirical ($\Lambda$CDM) cosmological simulation, using (extrapolated) radio luminosity functions for several different populations of sources. In this model, \frii{} radio sources dominate the counts above $\sim 200$~mJy, while below \fri{} sources dominate down to beyond our detection threshold. Figure~\ref{fig:bootes_dif_source_counts} shows an approximate agreement between the model and our observations, although the average power-law slope between 10 and 200~mJy of 0.83 is flatter than the value of 0.91 derived from our data.


For the models that are discussed here we find that all predicted source counts deviate from the observed source counts. The model source count slopes appear to be influenced mostly by the assumed contribution of \fri{} sources. On the other hand, the observational results from \citet{ishwara2010} also deviate from our result. More source counts from similar deep surveys at this frequency are needed to put stronger constraints on the source count models. Preliminary source counts from a deep GMRT 153~MHz survey three times the area of our current survey ($>1000$~sources; Intema et al., in preparation) match up closely with our current results. And future source counts from the ongoing GMRT 153~MHz sky survey (TGSS) should provide a more robust reference to compare the source count models and our results against.

\subsection{Spectral Indices}
\label{sec:bootes_spix}

\begin{figure}[!tp]
\begin{center}
\resizebox{\hsize}{!}{\includegraphics[angle=0]{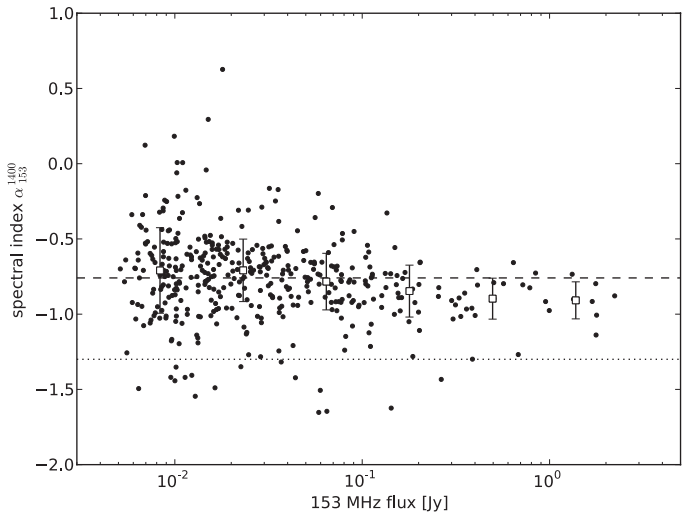}}
\caption{Spectral index between 1.4~GHz and 153~MHz for 417~sources, plotted as a function of the 153~MHz flux (black dots). The detection limit is almost fully determined by the 153~MHz survey due to the sensitivity of the 1.4~GHz observations. The median spectral index is $-0.76$ (dashed line). 16~sources have a spectral index below $-1.3$ (dotted line). The spectral indices are binned to emphasize the spectral flattening towards lower flux densities (white squares; see Table~\ref{tab:bootes_spix}).}
\label{fig:bootes_spectral}
\end{center}
\end{figure}

Because of the good match in resolution between the GMRT 153~MHz image and the deep WSRT 1.4~GHz image from \citet{devries2002}, we can accurately determine spectral indices over a decade in frequency. The survey depths at 153~MHz and 1.4~GHz are equal for sources with a spectral index of $-1.6$. Due to the high detection rate of 1.4~GHz sources at 153~MHz positions (Section~\ref{sec:bootes_compl}), we do an automated search for 1.4~GHz counterparts within $25\arcsec$ of the 153~MHz sources and ignore all sources for which we don't find counterparts. The spectral indices of 417~matched sources are plotted in Figure~\ref{fig:bootes_spectral}. We find a median spectral index of $-0.76$, which is similar to the median values of $-0.79$ \citep{cohen2004}, $-0.85$ \citep{ishwara2007}, $-0.82$ \citep{sirothia2009} and $-0.78$ \citep{ishwara2010} found for similar high-resolution, low-frequency surveys at 74 and 153~MHz. The small differences are most likely caused by differences in the completeness limits of the catalogs, as the median spectral index varies with flux density (see below).

Figure~\ref{fig:bootes_spectral} also plots the median spectral index in 6~logarithmic flux bins, which clearly highlights the flattening trend of the mean spectral index towards lower flux densities. The unique combination of our deep 153~MHz catalog and the very deep 1.4~GHz catalog makes that the median spectral index is unbiased down to the lowest 153~MHz flux densities. The median spectral index is $\sim -0.9$ for $\gtrsim 0.5$~Jy sources, and flattens to $\sim -0.7$ for $\lesssim 50$~mJy sources. A similar trend appears to be present in the spectral index distribution by \citet{ishwara2010} for sources between 153~MHz and various frequencies (610~MHz and higher), but may be biased by catalog flux limits at the higher frequencies. A similar flattening trend is also seen at both lower and higher frequencies, e.g., between 74~MHz and 1.4~GHz by \citet{cohen2004} and \citet{tasse2006}, between 1.4~GHz and 325~MHz by \citet{devries2002}, between 325~MHz and 1.4~GHz by \citet{zhang2003} and \citet{owen2009}, between 610~MHz and 1.4~GHz by \citet{bondi2007}, and between 1.4 and 5~GHz by \citet{prandoni2006}.

\ctable[botcap,center,
caption = {Median spectral index between 1.4~GHz and 153~MHz for 417~sources in 6~logarithmic (153~MHz) flux bins ranging from 5.0~mJy to 2.3~Jy.},
label = tab:bootes_spix
]{c c c}{
\tnote[a]{Median $\pm$ RMS around median.}
}{\FL
Flux bin center & Counts & Spectral index\tmark[a] \NN
{[Jy]} &  &  \ML
0.0083 & 133 & $-0.710\pm0.284$ \NN
0.0232 & 133 & $-0.709\pm0.208$ \NN
0.0643 &  84 & $-0.784\pm0.188$ \NN
0.1786 &  38 & $-0.847\pm0.173$ \NN
0.4966 &  19 & $-0.897\pm0.136$ \NN
1.3799 &  10 & $-0.908\pm0.123$ \LL
}


Recently, \citet{bornancini2010} found that, for a near-complete sample of radio galaxies, there is no evidence for spectral steepening or flattening due to redshifted curved radio spectra. In fact, most of their sources have straight power-law spectra from 74~MHz to 4.8~GHz. This suggests that the observed spectral flattening towards lower flux densities results dominantly from a correlation between source luminosity and spectral index ($P-\alpha$ correlation). This correlation is found to exist for \frii{} galaxies \citep[e.g.,][]{blundell1999}. According to the models by \citet{jackson2005} and \citet{wilman2008}, \frii{} galaxies are the dominant 153~MHz source population in our sample at higher flux levels ($\gtrsim 20-200$~mJy), which can explain part of the flattening observation. For lower flux levels, the flattening trend continues, which suggests that either \frii{} galaxies are still a significant source population, or the \fri{} galaxies dominate and also follow a $P-\alpha$ correlation.

From Figure~\ref{fig:bootes_spectral} we highlight a small group of 16~USS sources that have a spectral indices lower than $-1.3$. The lowest spectral index is $-1.65$, which indicates that the sample selection limit is indeed dominated by the 153~MHz survey depth. Table~\ref{tab:bootes_uss} lists these sources in decreasing flux order. Despite the relatively large uncertainty in 153~MHz flux density for the faintest sources, the spectral index is still relatively well constrained due to the large frequency span.

The fraction of USS sources is 3.8~percent (16 out of 417), which is mostly dependent on the survey detection limit at 153~MHz. It is not straightforward to compare this fraction with the few other 153~MHz surveys, as they use different high frequencies, with different detection limits at the high- and low-frequency end. Using the same criteria, \citet{sirothia2009} find a USS fraction of 3.7~percent (14 out of 374). Applying the same criteria to the source catalog by \citet{ishwara2010} yields a USS fraction of 3.0~percent (19 out of 639). These three fractions appear to have some consistency.

\ctable[botcap,center,star,
caption={153~MHz catalog selection of 16~USS sources with a spectral index $\alpha^{1400}_{153}$ below $-1.3$.},
label=tab:bootes_uss
]{c c c c c c}{
\tnote[a]{From \citet{devries2002}.}
\tnote[b]{Measured at 153~MHz.}
}{\FL
ID\tmark[a]  &  RA\tmark[b]  &  DEC\tmark[b]  &   153~MHz flux  &  1.4~GHz flux  &  $\alpha^{1400}_{153}$ \NN
 & & & [mJy] & [mJy] & \ML
J142656+352230  &  $14\hours26\mins56.46\secs$  &  $35\degs22\arcmin30.8\arcsec$  &  $264.2 \pm  26.6$  &  $11.31 \pm 0.46$  &  $-1.43 \pm 0.05$ \NN
J143506+350059  &  $14\hours35\mins06.89\secs$  &  $35\degs00\arcmin58.2\arcsec$  &  $142.8 \pm  14.5$  &  $ 4.02 \pm 0.16$  &  $-1.62 \pm 0.05$ \NN
J143118+351549  &  $14\hours31\mins18.29\secs$  &  $35\degs15\arcmin49.5\arcsec$  &  $ 64.8 \pm   6.9$  &  $ 1.74 \pm 0.08$  &  $-1.65 \pm 0.05$ \NN
J143500+342531  &  $14\hours35\mins00.98\secs$  &  $34\degs25\arcmin30.2\arcsec$  &  $ 59.7 \pm   6.2$  &  $ 2.18 \pm 0.09$  &  $-1.51 \pm 0.05$ \NN
J143520+345950  &  $14\hours35\mins20.51\secs$  &  $34\degs59\arcmin49.1\arcsec$  &  $ 58.6 \pm   6.2$  &  $ 1.55 \pm 0.07$  &  $-1.65 \pm 0.05$ \NN
J143815+344428  &  $14\hours38\mins15.28\secs$  &  $34\degs44\arcmin29.8\arcsec$  &  $ 44.0 \pm   4.8$  &  $ 1.93 \pm 0.09$  &  $-1.42 \pm 0.05$ \NN
J143331+341012  &  $14\hours33\mins31.84\secs$  &  $34\degs10\arcmin12.9\arcsec$  &  $ 37.0 \pm   4.1$  &  $ 2.04 \pm 0.09$  &  $-1.32 \pm 0.05$ \NN
J142631+341557  &  $14\hours26\mins31.69\secs$  &  $34\degs16\arcmin00.9\arcsec$  &  $ 22.5 \pm   3.6$  &  $ 1.16 \pm 0.07$  &  $-1.35 \pm 0.08$ \NN
J142954+343516  &  $14\hours29\mins53.90\secs$  &  $34\degs35\arcmin18.8\arcsec$  &  $ 16.4 \pm   2.4$  &  $ 0.62 \pm 0.04$  &  $-1.49 \pm 0.07$ \NN
J142724+334714  &  $14\hours27\mins24.77\secs$  &  $33\degs47\arcmin18.5\arcsec$  &  $ 12.8 \pm   2.8$  &  $ 0.43 \pm 0.04$  &  $-1.55 \pm 0.11$ \NN
J143538+335347  &  $14\hours35\mins38.76\secs$  &  $33\degs53\arcmin44.2\arcsec$  &  $ 12.3 \pm   2.2$  &  $ 0.56 \pm 0.05$  &  $-1.41 \pm 0.09$ \NN
J143310+333131  &  $14\hours33\mins10.42\secs$  &  $33\degs31\arcmin27.7\arcsec$  &  $ 11.3 \pm   2.2$  &  $ 0.50 \pm 0.04$  &  $-1.42 \pm 0.10$ \NN
J143230+343449  &  $14\hours32\mins30.47\secs$  &  $34\degs34\arcmin49.5\arcsec$  &  $ 10.1 \pm   1.8$  &  $ 0.52 \pm 0.04$  &  $-1.35 \pm 0.09$ \NN
J142719+352326  &  $14\hours27\mins19.32\secs$  &  $35\degs23\arcmin29.2\arcsec$  &  $ 10.0 \pm   3.4$  &  $ 0.42 \pm 0.06$  &  $-1.44 \pm 0.17$ \NN
J143700+335920  &  $14\hours37\mins00.74\secs$  &  $33\degs59\arcmin20.2\arcsec$  &  $  9.5 \pm   2.5$  &  $ 0.42 \pm 0.04$  &  $-1.42 \pm 0.13$ \NN
J143249+343915  &  $14\hours32\mins49.12\secs$  &  $34\degs39\arcmin14.0\arcsec$  &  $  6.4 \pm   1.8$  &  $ 0.24 \pm 0.04$  &  $-1.49 \pm 0.15$ \LL
}

The angular distribution of the USS sources is quite peculiar. There are 6~sources that form 3~pairs within $6\arcmin$ of each other. These pairs, together with the remaining 10~single sources, appear to be randomly distributed across the FoV. Visual inspection of the pairs in the images at 153~MHz and 1.4~GHz did not reveal any obvious artifacts in the background near these sources, which makes it less likely that these sources are fake detections. There is no visual evidence in the radio maps that the pair components are physically connected. Through Monte-Carlo simulations we determined that, out of 16~sources with random positions over a 1.5~degree radius field (which is approximately the 1.4~GHz field radius), the chance of finding 3~pairs within $6\arcmin$ is $1.4 \pm 0.4$~percent. Therefore, it is unlikely that these pairs appeared by chance. Further investigation is needed to establish the true nature of these pairs.

\citet{croft2008} examine 4~candidate HzRGs (which they labelled A, B, C and E) based on their steep ($\leq -0.87$) spectral index between 1.4~GHz and 325~MHz. As their candidate sources are also present in the 153~MHz catalog, we complement their data with our new spectral index measurements in Table~\ref{tab:bootes_croft}. Both sources~A and E appear to have fairly straight power-law spectra down to 153~MHz, while sources~B and C appear to undergo considerable spectral flattening. Based on our selection criteria for USS sources, only source~A would be considered a candidate HzRG.

\ctable[botcap,center,star,
caption={Spectral indices between 1.4~GHz and 153~MHz for four candidate HzRGs by \citet{croft2008}.},
label=tab:bootes_croft
]{l c c c c c}{
\tnote[a]{As measured at 1.4~GHz; \citet{devries2002}.}
\tnote[b]{The spectral index between 325~MHz and 1.4~GHz as measured by \citet{croft2008}.}
\tnote[c]{The spectral index between 153~MHz and 1.4~GHz as measured in this work.}
\tnote[d]{Photometric redshift, as measured by \citet{croft2008}.}
}{\FL
ID & RA\tmark[a] & DEC\tmark[a] & $\alpha^{1400}_{325}$ (\tmark[b]) & $\alpha^{1400}_{153}$ (\tmark[c]) & $z^{}_\mathrm{phot}$\tmark[d] \ML
A  & $14\hours26\mins31.75\secs$ & $+34\degs15\arcmin57.5\arcsec$ & -1.48 & $-1.35 \pm 0.08$ & 4.97 \NN
B  & $14\hours26\mins47.87\secs$ & $+34\degs58\arcmin51.0\arcsec$ & -0.89 & $-0.56 \pm 0.06$ & 3.76 \NN
C  & $14\hours27\mins41.84\secs$ & $+34\degs23\arcmin24.7\arcsec$ & -0.98 & $-0.44 \pm 0.13$ & 1.21 \NN
E  & $14\hours32\mins58.44\secs$ & $+34\degs20\arcmin55.4\arcsec$ & -0.87 & $-0.95 \pm 0.05$ & 4.65 \LL
}

\subsection{Identification Fraction of Radio Sources versus Spectral Index}
\label{sec:bootes_id_fraction}

We investigate the NIR $K$-band identification fraction of radio sources as function of spectral index \citep[e.g.,][]{wieringa1991}. This links together two known correlations, namely the correlation between spectral index and redshift \citep{tielens1979,blumenthal1979} and the correlation between $K$-band magnitude and redshift \citep[$K-z$ correlation; e.g.,][]{willott2003,rocca2004}. We remark that the correlation between spectral index and redshift is not a tight correlation; for a given spectral index there is a range of possible redshifts for a radio source \citep[e.g.,][]{miley2008}. But on average, steep spectrum sources are located at higher redshifts, and are consequently (on average) more difficult to detect in optical and NIR bands. Using $K$-band has an advantage over (also available) shorter wavelength bands as this band suffers the least from extinction.

We have identified possible optical counterparts of the radio sources using the FLAMEX $K^{}_{S}$-band catalogue \citep{elston2006}. This survey covers 7.1~square degrees within the \bootes{} field. For the NIR identification we use the likelihood ratio technique described by \citet{sutherland1992}. This allows us to obtain an association probability for each NIR counterpart, taking into account the NIR magnitudes of the possible counterparts. Before the cross identification, we removed all radio sources located outside the coverage area of the $K^{}_{S}$-band images, or located within 20~pixels of the edge or other blanked pixels within individual $K^{}_{S}$-band frames.

Following \citet{sutherland1992} and \citet{tasse2008}, the probability that a NIR counterpart with magnitude $m$ is the true NIR counterpart of the radio source is given by the likelihood ratio
\begin{equation}
\mathrm{LR}(r,m) = \frac{ \theta(<m) \exp{(-r^{2}_{}/2)} }{ 2 \pi \, \sigma_{\alpha} \, \sigma^{}_{\delta} \, \rho(<m)},
\end{equation}
with $m$ the $K^{}_{S}$-band magnitude of the NIR candidate, $\theta(<m)$ the a priori probability that the radio source has a NIR counterpart with a magnitude brighter than $m$, and $\rho(<m)$ the surface number density of NIR sources with a magnitude smaller than $m$. $\sigma^{2}_{\delta}$ and $\sigma^{2}_{\alpha}$ are the quadratic sums of the uncertainties in the radio and NIR positions in right ascension and declination, respectively. For the radio source positions and uncertainties we have taken the values from the 1.4~GHz WSRT catalogue, as the astrometric precision is better than the 153~MHz GMRT data. For the FLAMEX survey we have adopted $0.3\arcsec$ for the position uncertainty (this is half the CCD pixel size, as source positions are truncated to pixel positions). The uncertainty-normalized angular distance is given by the parameter $r = \sqrt{(\Delta^{}_{\alpha} / \sigma^{}_{\alpha})^2  + (\Delta^{}_{\delta} / \sigma^{}_{\delta})^{2}_{}}$, with $\Delta$ the positional difference in either $\alpha$ or $\delta$ between the possible NIR counterpart and radio source.

The probability $P(i)$ that candidate $i$ is the true NIR counterpart is
\begin{equation}
P(i) = \frac{ \mathrm{LR}^{}_{i}(r,m) }{ \sum_{j} \mathrm{LR}^{}_{j}(r,m) + [ 1 - \theta(m^{}_\mathrm{lim}) ] },
\end{equation}
with $j$ running over all possible NIR counterparts. $\theta(m^{}_\mathrm{lim})$ is the fraction of radio sources having a NIR counterpart at the magnitude limit of the NIR survey. We found $\theta(m_\mathrm{lim})$ to be 0.62, where $m_\mathrm{lim} = 19.2$. The values for $\rho(<m)$ (surface number density of NIR sources) were estimated from the data itself using bins of $0.1$~magnitude across the full survey area. The a-priori identification fraction $\theta(<m)$ was also estimated from the data itself, using the technique described by \citet{ciliegi2003}. This involves counting the surface number density of NIR sources around the radio sources as function of magnitude. This distribution is compared to a distribution of background objects, the overdensity of NIR sources around radio sources gives an estimate of $\theta(<m)$.

\begin{figure}[!tp]
\begin{center}
\resizebox{\hsize}{!}{\includegraphics[angle=0]{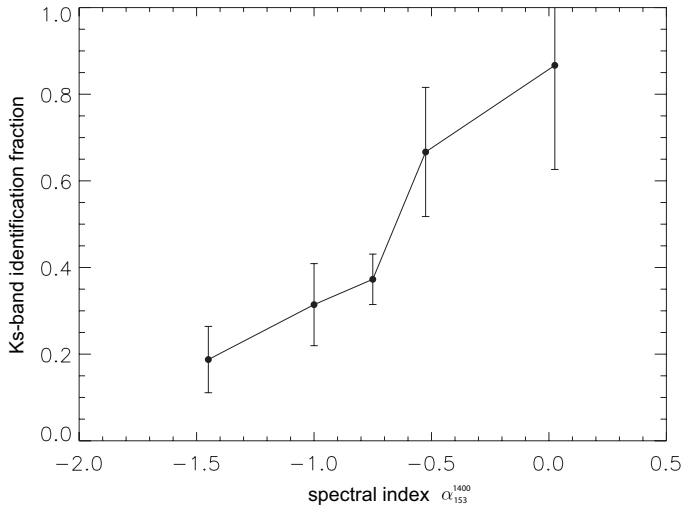}}
\caption{Identification fraction of radio sources in the near-infrared $K^{}_{S}$-band as a function of spectral index between 1.4~GHz and 153~MHz.}
\label{fig:bootes_identification}
\end{center}
\end{figure}

We selected 368~radio sources that are present within both the 1.4~GHz and 153~MHz catalogs (matched within $6\arcsec$) and have a simple morphology (fitted with a single gaussian). We have computed the likelihood ratio of all NIR counterparts located within $20\arcsec$ from the radio position. We have defined a radio source to have a NIR counterpart if $\sum_{j} \mathrm{LR}^{}_{j}(r,m) > 0.75$. The results are shown in Figure~\ref{fig:bootes_identification}. The identification fraction is $\sim 70$~percent for $\alpha^{1400}_{153} > -0.7$, while for $\alpha^{1400}_{153} < -0.7$ the identification fraction drops to about 30~percent. This result reproduces the expected (and previously observed) correlation between spectral index and $K$-band identification fraction \citep[e.g.,][]{wieringa1991}.

\section{Conclusions and Future Plans}
\label{sec:bootes_concl}

We have presented the results from a deep ($\sim 1.0$~mJy central noise), high-resolution ($26\arcsec \times 22\arcsec$) radio survey at 153~MHz, covering the full NOAO \bootes{} field and beyond. This 11.3~square degree survey is amongst the deepest surveys at this frequency to date. We produced a catalog of 598~sources detected at 5~times the local noise level, with source flux densities ranging from 3.9~Jy down to 5.1~mJy. We estimate the catalog to be $\sim 70$~percent complete at the $5.1$~mJy flux limit, $\sim 92$~percent for $> 10$~mJy sources and $> 99$~percent for $> 20$~mJy sources, and less than $1$~percent contaminated by false detections. The on-source dynamic range (with noise measured near the source) is limited to $\sim 600$, while off-source (noise measured in central part of the image, away from bright sources) this rises to $>1000$. We expect that residual RFI and residual direction-dependent calibration errors prevents reaching the thermal noise level of 0.2-0.3~\mjybeam{}.

The 153~MHz catalog presented in Section~\ref{sec:bootes_bdsm} allows for a detailed study of source populations in a relatively unexplored flux range. We have analyzed the source counts and spectral index distributions for our survey. From this analysis we draw the following conclusions: \\
(i) The Euclidean-normalized differential source counts, determined over a flux range from 10~mJy to 1~Jy, are well approximated by a single power-law slope of 0.91. The inconsistency with model source counts by \citet{jackson2005} and \citet{wilman2008} seems due to uncertainty in the predicted contribution of \fri{} galaxies. Our survey is not deep enough to detect the flattening at the lowest flux densities as seen at higher frequencies (Section~\ref{sec:bootes_dsc}). \\
(ii) The spectral indices of a subset of 417~sources between 153~MHz and 1.4~GHz, limited predominantly by the 153~MHz sensitivity, have a median spectral index of $-0.76$. We confirm the flattening trend in the median spectral index towards lower 153~MHz flux densities, as seen at higher frequencies (Section~\ref{sec:bootes_spix}). \\
(iii) The detection fraction of USS sources, having a spectral index below -1.3, 
is 3.8~percent, which is comparable to the USS fractions of two comparable 153~MHz surveys. Six of the 16~USS sources, apparently physically unrelated, are found in pairs $< 6\arcmin$ apart, which is an unlikely spatial distribution to appear at random (Section~\ref{sec:bootes_spix}). \\
(iv) The NIR $K_S$-band identification fraction of a subset of 368~sources drops from $\sim 0.7$ for spectral indices $> -0.7$ to $\sim 0.3$ for spectral indices $< -0.7$, reproducing a previously observed correlation \citep[e.g.,][]{wieringa1991} which links together the known correlations between $K$-band magnitude and redshift, and between spectral index and redshift \citep{tielens1979,blumenthal1979,willott2003,rocca2004} (Section~\ref{sec:bootes_id_fraction}).

We plan to continue our analysis of the \bootes{} 153~MHz source survey by comparing it with many other available catalogs at various spectral bands. A high priority task will be to study the properties of the 16~USS sources, to determine if these objects are HzRGs, derive estimates of their redshifts and search the surrounding area for galaxies at similar redshift. This approach has been succesful for the identification and study of galaxy cluster formation \citep[e.g.,][]{rottgering1994,venemans2002}.

The observations presented here are part of a larger survey with six additional, partly overlapping flanking fields observed with GMRT at the same frequency, covering a total survey area of $\sim 40$~square degrees. This same area is also covered by extended WSRT observations using 8~bands between 115--165~MHz. We will combine these observations with the observations presented here to produce a combined high- and low-resolution catalog at $\sim 153$~MHz to further facilitate the study of the low-frequency sky, and in particular to facilitate the further search for USS radio sources.

\begin{acknowledgements}

The authors thank Jim Condon, Ed Fomalont and Bill Cotton for useful discussions, and the anonymous referee for providing useful comments. We would also like to thank the staff of the GMRT that made these observations possible. GMRT is run by the National Centre for Radio Astrophysics of the Tata Institute of Fundamental Research. This study made use of online available maps and catalogs from the WSRT \bootes{} Deep Field survey \citep{devries2002} and the FLAMINGOS Extragalactic Survey \citep{elston2006}. We are grateful to Niruj Mohan for providing support for the source extraction tool BDSM. HTI acknowledges a grant from the Netherlands Research School for Astronomy (NOVA). This work was partly supported by funding associated with an Academy Professorship of the Royal Netherlands Academy of Arts and Sciences (KNAW), and by the National Radio Astronomy Observatory, which is operated by Associated Universities, Inc., under cooperative agreement with the National Science Foundation.

\end{acknowledgements}

\bibliographystyle{aa}
\bibliography{14253}

\appendix
\onecolumn

\section{Selection of 153~MHz Radio Source Images}
\label{sec:bootes_images}

\begin{figure}[!h]
\begin{center}
\resizebox{\hsize}{!}{
\includegraphics[angle=0]{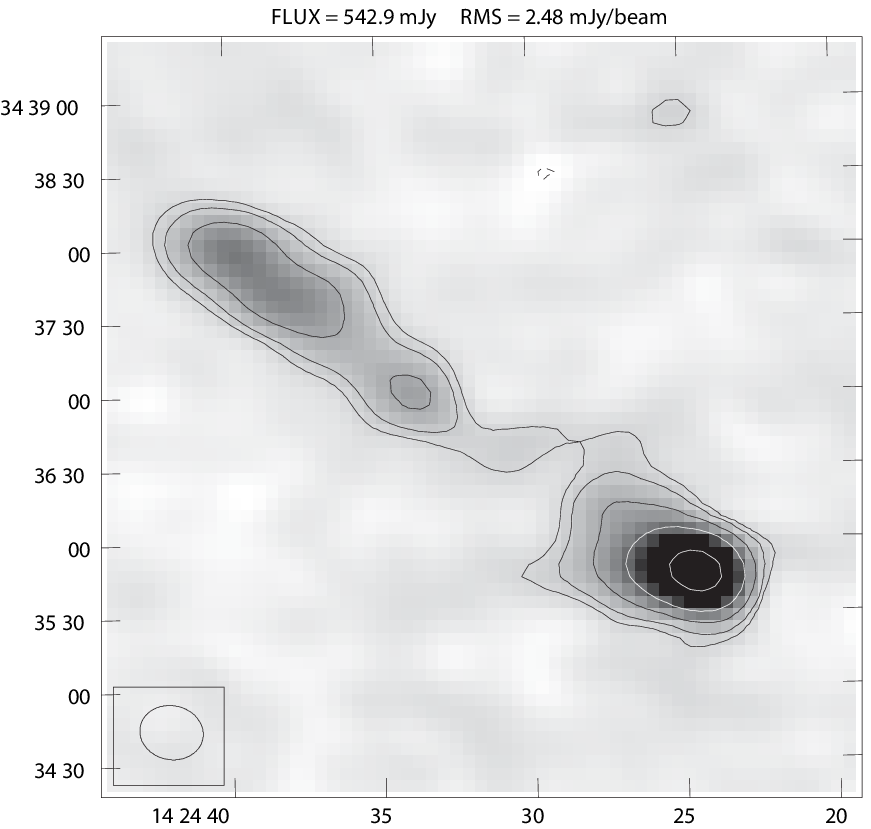}
\includegraphics[angle=0]{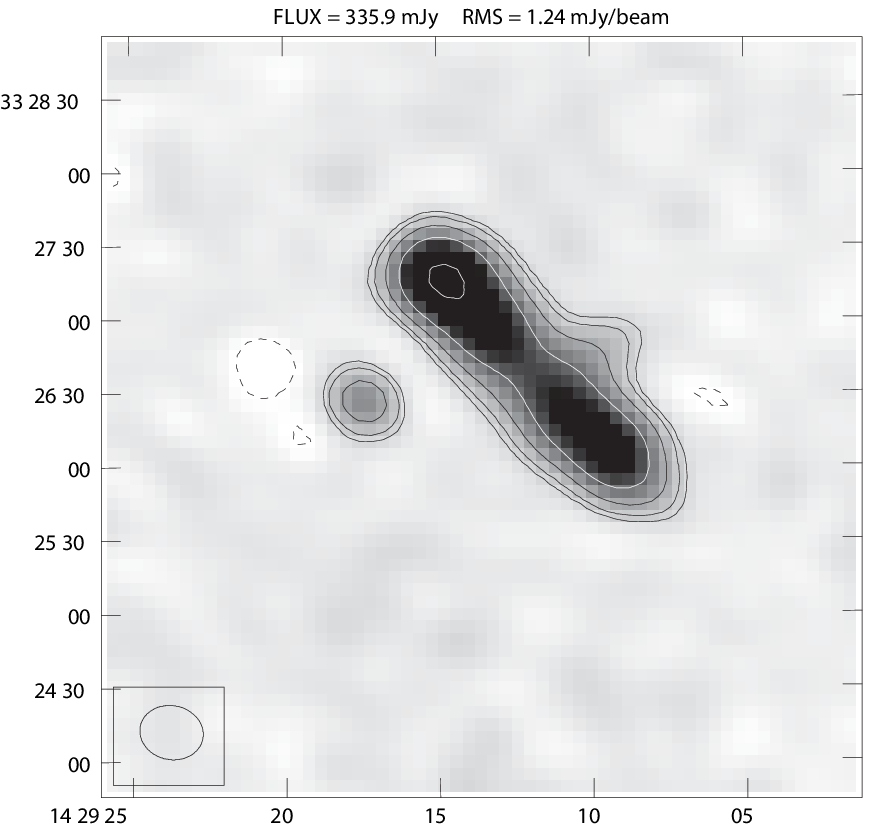}
\includegraphics[angle=0]{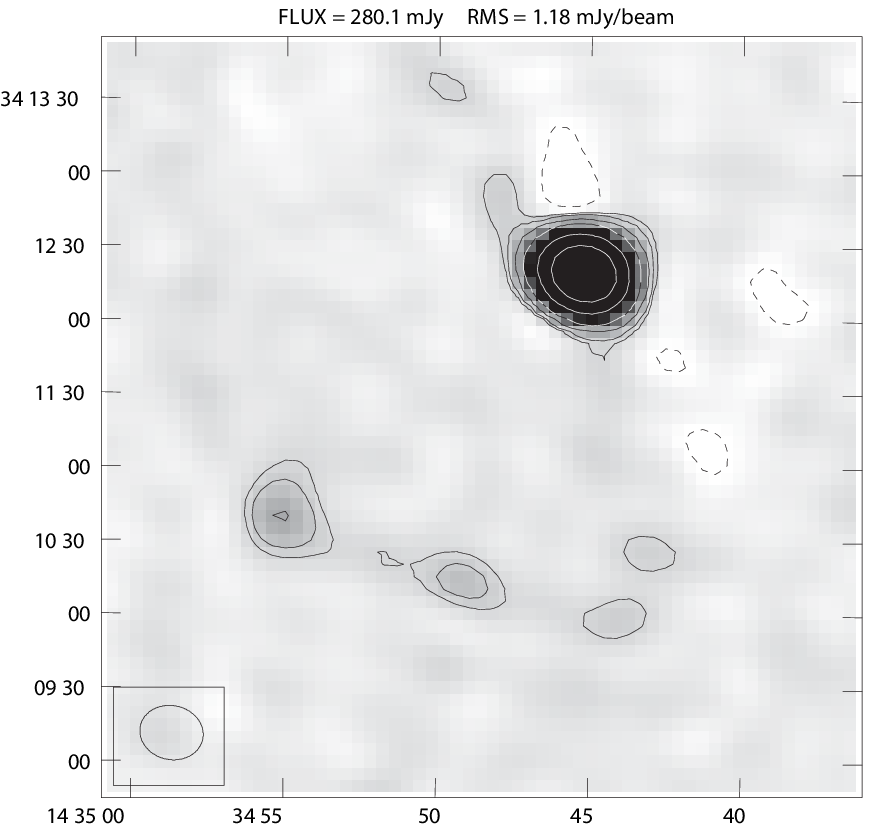}
\includegraphics[angle=0]{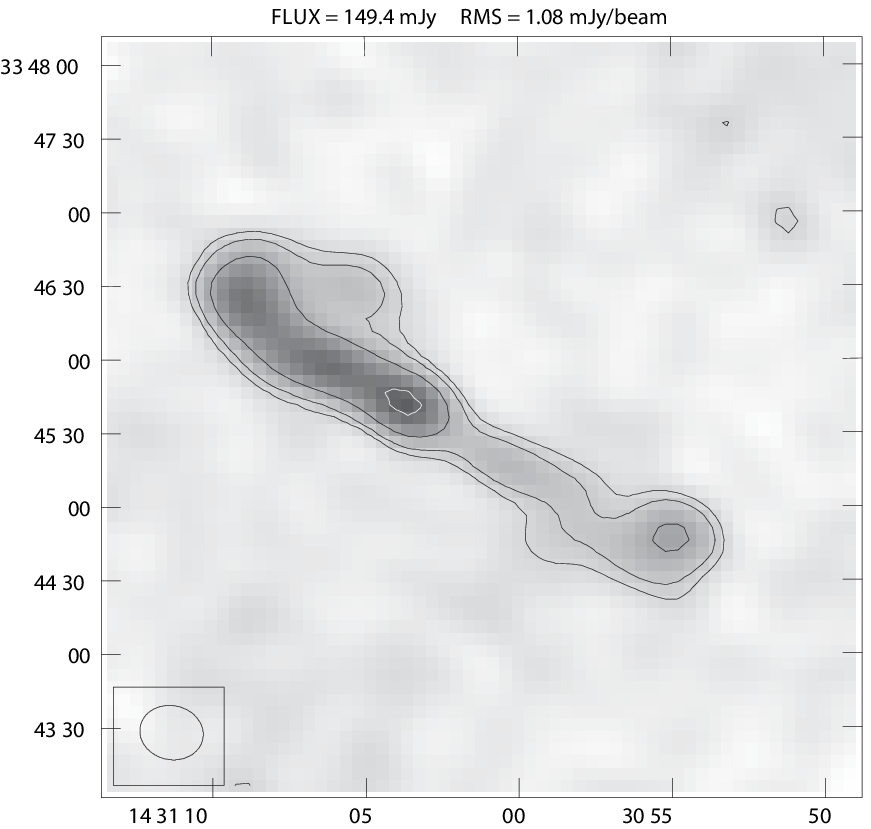}
} \\
\resizebox{\hsize}{!}{
\includegraphics[angle=0]{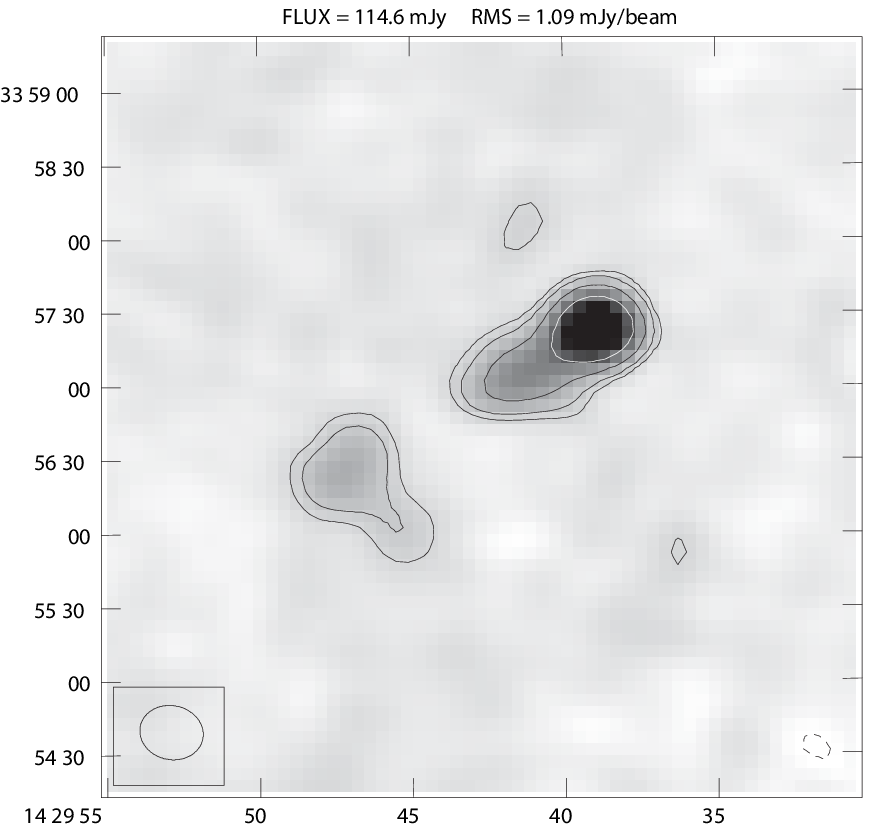}
\includegraphics[angle=0]{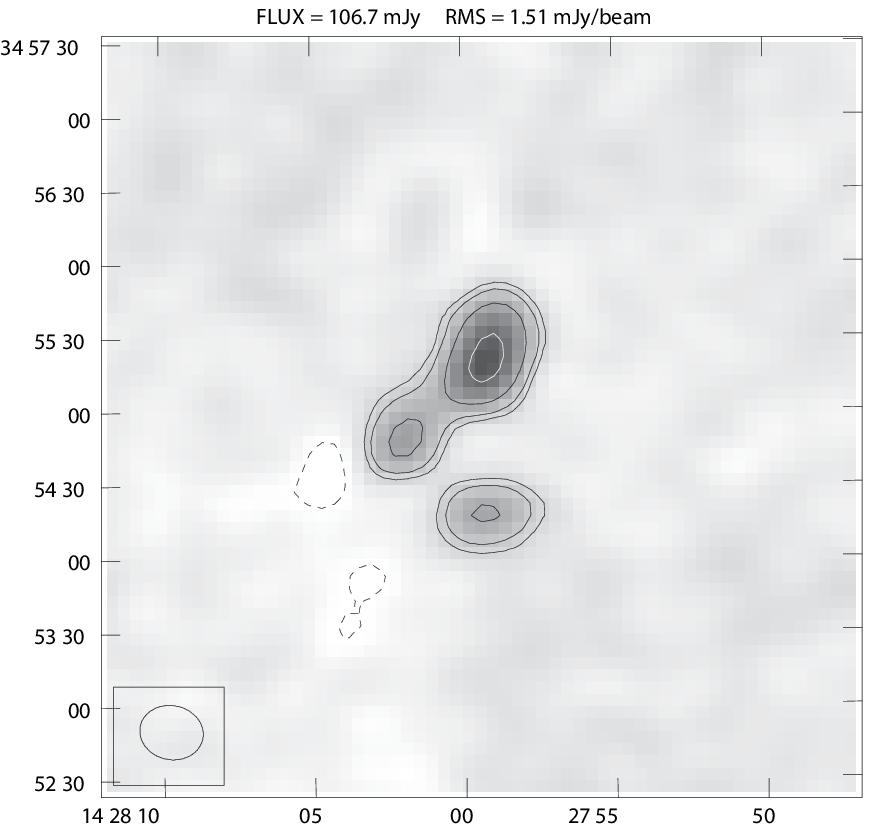}
\includegraphics[angle=0]{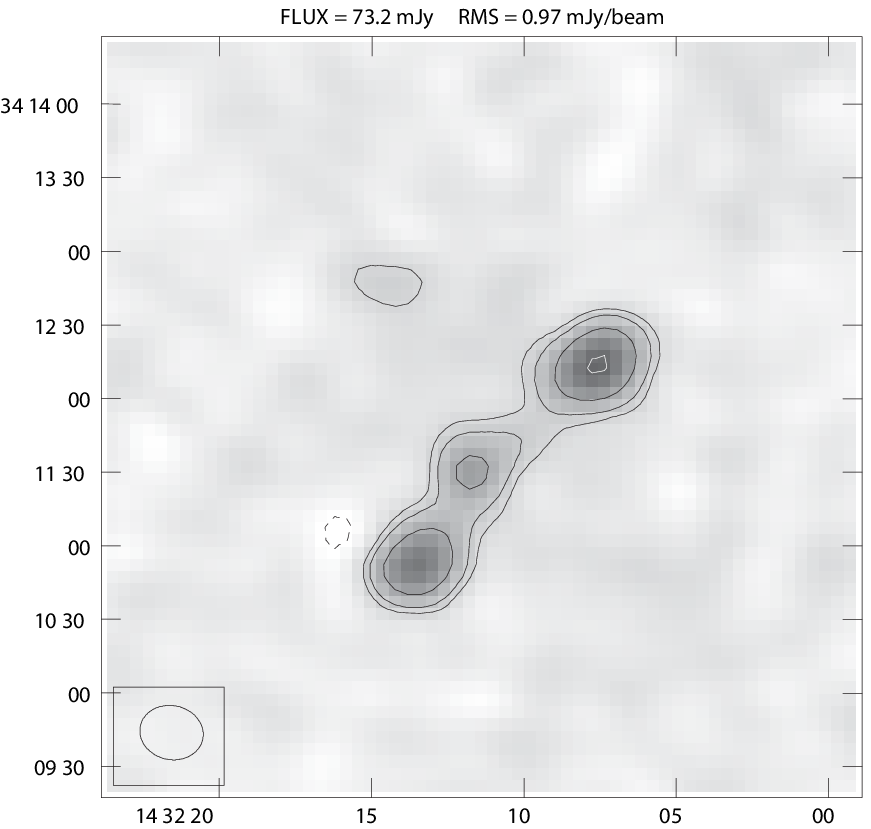}
\includegraphics[angle=0]{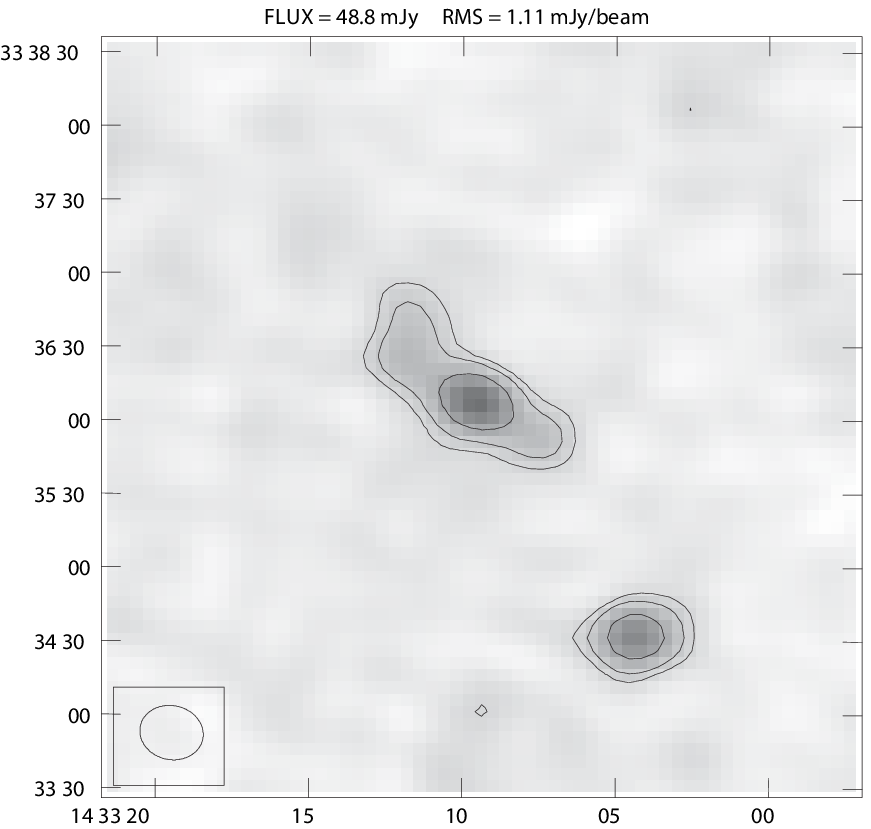}
} \\
\resizebox{\hsize}{!}{
\includegraphics[angle=0]{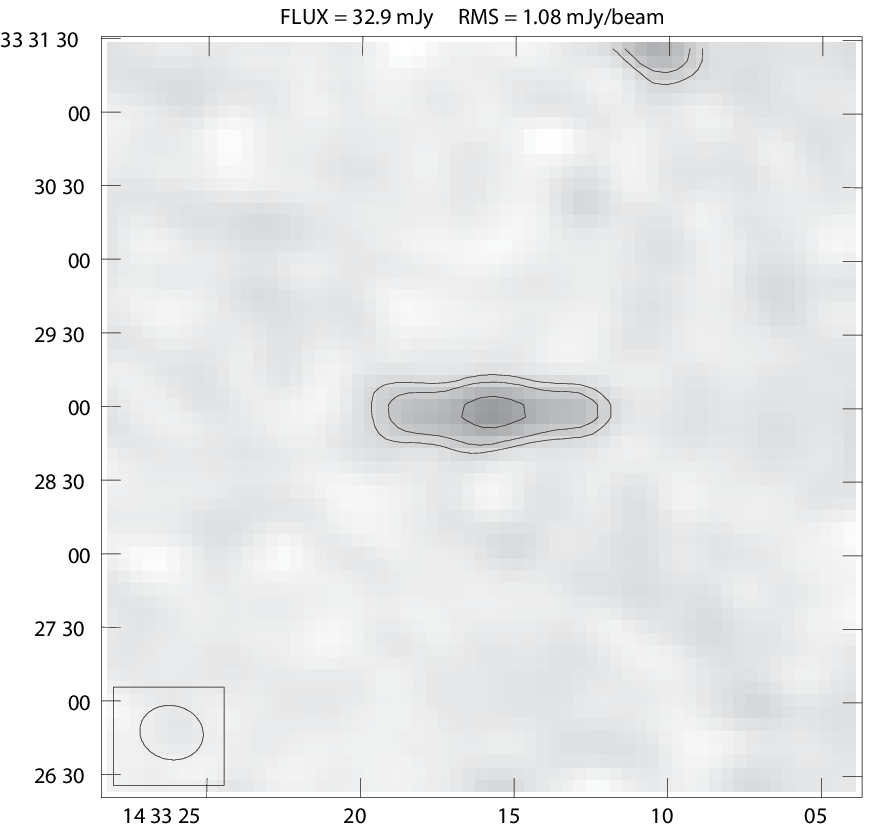}
\includegraphics[angle=0]{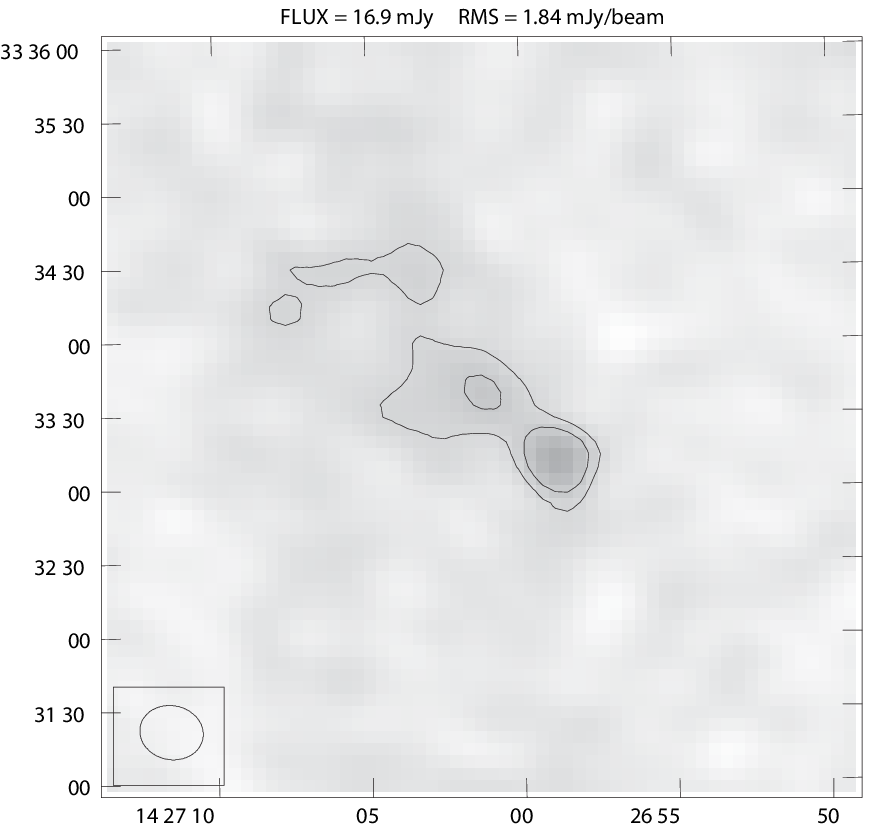}
\includegraphics[angle=0]{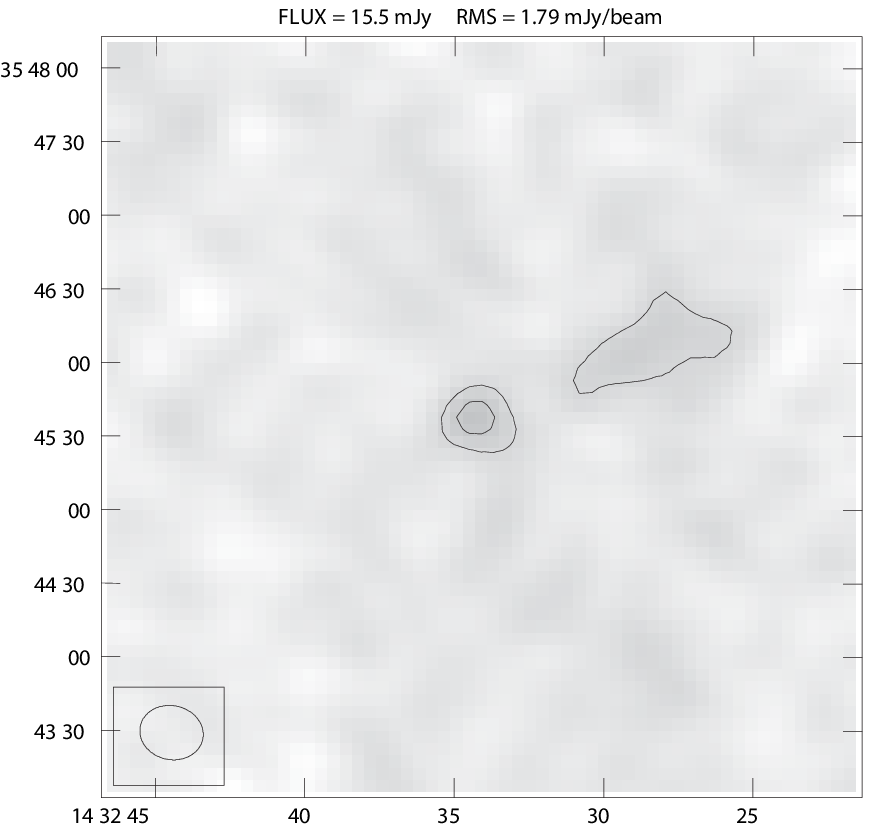}
\includegraphics[angle=0]{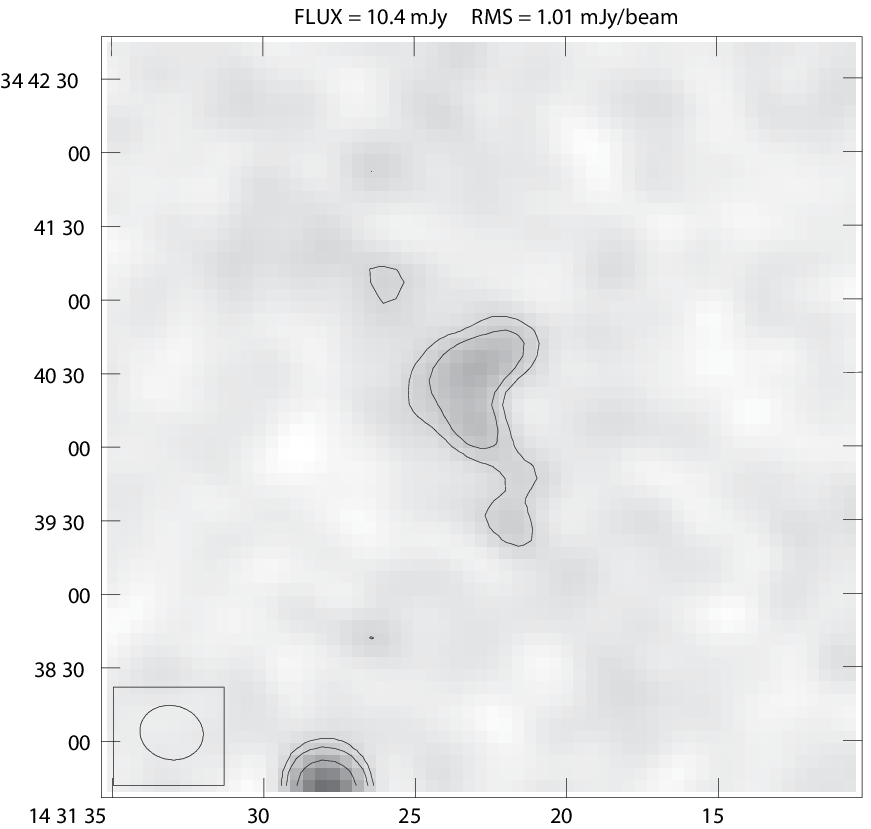}
}
\caption{Grey-scale images of a selection of resolved 153~MHz sources with a complex morphology (i.e., not a point source or double point source morphology). Horizontal and vertical axes are RA and DEC, respectively. For each source, the total flux and local RMS noise are specified above the $5\arcmin \times 5\arcmin$ image. Contours are drawn at $[-3,3,5,10,20,50,100]$ times the local noise. The grey-scale ranges from $-3$ to 30 times the local noise.}
\label{fig:bootes_images}
\end{center}
\end{figure}

\end{document}